\documentclass[twocolumn,twocolappendix]{aastex63}
\usepackage{nicefrac}
\usepackage{amsmath,bm}
\usepackage{float}
\usepackage{tabularx}
\usepackage{mathtools}
\usepackage[T1]{fontenc}

\newcommand{\ud}{\mathrm{d}}

\newcommand{\dd}[2]{\dfrac{\ud #1}{\ud #2}}
\newcommand{\bhspin}{a_\star}

\newcommand{\bk}{{\boldsymbol{k}}}
\renewcommand{\bv}{{\boldsymbol{v}}}
\newcommand{\rA}{{\rm A}}

\newcommand{\alf}{{Alfv\'en~}}
\def\iharm3d{{\tt{}iharm3D}}

\defcitealias{PaperI}{EHTC~I}
\defcitealias{PaperII}{EHTC~II}
\defcitealias{PaperIII}{EHTC~III}
\defcitealias{PaperIV}{EHTC~IV}
\defcitealias{PaperV}{EHTC~V}
\defcitealias{PaperVI}{EHTC~VI}

\shortauthors{Wong et al.}

\begin{document}

\title{The Jet--Disk Boundary Layer in Black Hole Accretion}

\correspondingauthor{George~N.~Wong}
\email{gnwong2@illinois.edu}

\author[0000-0001-6952-2147]{George~N.~Wong}
\affil{Illinois Center for Advanced Studies of the Universe, Department of Physics, \\University of Illinois, 1110 West Green Street, Urbana, IL 61801, USA}
\affil{CCS-2, Los Alamos National Laboratory, P.O.~Box 1663, Los Alamos, NM 87545, USA}
\author[0000-0003-0510-5170]{Yufeng Du}
\affil{Illinois Center for Advanced Studies of the Universe, Department of Physics, \\University of Illinois, 1110 West Green Street, Urbana, IL 61801, USA}
\affil{Division of Physics, Mathematics and Astronomy, California Institute of Technology, Pasadena, CA 91125, USA}
\author[0000-0002-0393-7734]{Ben~S.~Prather}
\affil{Illinois Center for Advanced Studies of the Universe, Department of Physics, \\University of Illinois, 1110 West Green Street, Urbana, IL 61801, USA}
\affil{CCS-2, Los Alamos National Laboratory, P.O.~Box 1663, Los Alamos, NM 87545, USA}
\author[0000-0001-7451-8935]{Charles~F.~Gammie}
\affil{Illinois Center for Advanced Studies of the Universe, Department of Physics, \\University of Illinois, 1110 West Green Street, Urbana, IL 61801, USA}
\affil{Department of Astronomy, University of Illinois, 1002 West Green Street, Urbana, IL 61801, USA}

\begin{abstract}

Magnetic fields lines are trapped in black hole event horizons by accreting plasma.  If the trapped field lines are lightly loaded with plasma, then their motion is controlled by their footpoints on the horizon and thus by the spin of the black hole.  In this paper, we investigate the boundary layer between lightly loaded polar field lines and a dense, equatorial accretion flow.  We present an analytic model for aligned prograde and retrograde accretion systems and argue that there is significant shear across this ``jet--disk boundary'' at most radii for all black hole spins. Specializing to retrograde aligned accretion, where the model predicts the strongest shear, we show numerically that the jet--disk boundary is unstable.  The resulting mixing layer episodically loads plasma onto trapped field lines where it is heated, forced to rotate with the hole, and permitted to escape outward into the jet.  In one case we follow the mass loading in detail using Lagrangian tracer particles and find a time-averaged mass-loading rate $\sim 0.01\, \dot{M}$. 

\end{abstract}


\section{Introduction}

According to Alfv\'en's theorem, magnetic fields lines are frozen into highly conducting plasmas and are advected with the plasmas as they move under the influence of external forces. This freeze-in effect operates near black holes when the accreting plasma falls onto the hole, and thus it is natural for a black hole to have field lines that thread its event horizon.  If the horizon-threading field lines are open and lightly loaded with plasma so that the local magnetization\footnote{Here, $B$ is the strength of the magnetic field, $\rho$ is the rest-mass density of the plasma, and $c$ is the speed of light. In this paper, we use Lorentz--Heaviside units for electromagnetic quantities.} is much larger than unity
\begin{equation}
    \sigma \equiv \frac{B^2}{\rho c^2} \gg 1
\end{equation}
in the region close to the horizon, then their motion is controlled by gravity, and they are forced to rotate if the black hole has nonzero spin. 

Forced rotation of field lines was first studied by \cite[hereafter BZ]{blandford77} who solved a force-free magnetosphere model in the limit that the black hole dimensionless spin $\bhspin \equiv Jc/GM^2 \ll 1$ (here $J \equiv$ spin angular momentum, $M \equiv$ mass).  BZ found that the field behaves as if it were anchored in a star rotating with frequency
\begin{equation}
    \Omega_F \equiv \frac{1}{2}\Omega_H = \frac{\bhspin}{8} \frac{c^3}{G M} +  O(\bhspin^3),
\end{equation}
where $\Omega_H$ is the rotation frequency of the event horizon. Field line rotation produces an outward-directed energy current at the horizon.  In the force-free limit this is known as the BZ effect, whereas
if the field lines are more heavily loaded it is also sometimes called the magnetohydrodynamic (MHD) Penrose process \citep{takahashi1990}.  The BZ effect is a favored mechanism for powering extragalactic radio jets.

In recent decades, numerical general relativistic magnetohydrodynamics (GRMHD) simulations have been used to study black hole accretion and the BZ mechanism (see \citet{Davis2020} and \citet{Komissarov2021} for reviews).
In GRMHD models with a trapped magnetic flux $\Phi$, a low density region forms around an axis parallel to the accretion flow angular momentum vector as plasma falls down the field lines into the hole or is expelled to larger radius.  This low density region, with $\sigma \gg 1$, contains horizon-threading field lines moving with rotation frequency $\Omega_F$ and an associated, outward-directed energy current \citep[Poynting flux;][]{mckinney04}.  In what follows we will refer to this region as the {\em jet}. It is difficult for numerical codes to robustly evolve parts of the simulation domain with low density and high $\sigma$, like in the jet, so semi-analytic magnetosphere models are often invoked to study these regions \citep[see, e.g.,][]{Ogihara2021}.

The jet is bounded by an accretion flow that pins magnetic flux in the hole. We will refer to the accretion flow as a {\em disk}, although it may have sub-Keplerian rotation.  At the boundary layer between the jet and the disk, the density contrast is large. The plasma velocity can also change dramatically, with maximal shear occurring when the black hole and disk rotate in opposite directions (a \emph{retrograde} disk).

The jet--disk boundary layer has large shear and strong currents.  It can suffer instabilities that lead to mass loading onto the jet's open field lines.  It may also be an important particle acceleration site (see the reviews of (see the reviews of \citealt{Ostrowski1999, Rieger2019} for particle acceleration in relativistic shear layers).  This paper considers the jet--disk boundary layer in the relativistic regime, within $\sim 20\;GM/c^2$ of the event horizon.  

In Section~\ref{sec:estimates} we provide simple estimates for shear at the jet--disk boundary layer.   In Section~\ref{sec:grmhd} we describe the GRMHD simulations we use to study the jet--disk boundary layer, and in Section~\ref{sec:results}, we explore the dynamics of the boundary layer by using tracer particles to both analyze the flow of matter through state space and investigate mass loading into the jet.  Along the way we discuss the disk structure for retrograde accretion.  In Section~\ref{sec:discussion} we consider model limitations, convergence, and possible extensions.  Section~\ref{sec:summary} provides a summary and a guide to the main results.

\section{Scaling and Estimates}
\label{sec:estimates}

We now define the physical parameters that describe accretion systems, identify their ranges for the systems we consider, and provide an analytic estimate for flow dynamics at the jet--disk boundary layer.

\subsection{Parameters}

We consider radiatively inefficient accretion flows (RIAFs; \citealt{Reynolds96}) where radiative cooling is negligible, motivated by EHT observations of M87* and Sgr A*, which have accretion rate $\dot{m} \equiv \dot{M}/\dot{M}_{\mathrm{Edd}} \ll 1$ ($\dot{M}_{\mathrm{Edd}}$ is the Eddington accretion rate) and are therefore near or in this regime.  RIAFs are geometrically thick disks, with ratio of scale height $H$ to local radius $R$ of order $1$.

In general, the angular momentum of accreting matter far from the horizon may be tilted with respect to the black hole's spin angular momentum.  Although there are plausible scenarios that produce zero tilt, there is at present no way of rejecting models with strong or even maximal ($180$ degree) tilt.  In this paper we restrict attention to systems where the orbital angular momentum of the accreting plasma is parallel or anti-parallel to the black hole spin vector (prograde or zero tilt and retrograde or maximal tilt, respectively).  Disks with intermediate tilt are the subject of ongoing study \citep{Fragile2007, McKinney2013, moralesteixeira2014, Liska2018, White2019}.  

In addition to $\bhspin$, $\dot{m}$, and tilt, black hole accretion flows are characterized by $\Phi$, the trapped magnetic flux measured through the contour formed by the black hole's equator.  Accretion of flux with a consistent sign eventually increases $|\Phi|$ until the accumulated magnetic flux is large enough that magnetic pressure $B^2 \sim (\Phi/(G M/c^2)^2)^2$ balances accretion ram pressure $\rho c^2$.  Since $\dot{M} \sim \rho c (G M/c^2)^2$, when the dimensionless flux $\phi \equiv \Phi/\sqrt{G^2 M^2 \dot{M}/c^3}$ approaches a critical value $\phi_c \sim 15$ (\citealt{Tchekhovskoy2011maddefn}, but we use the normalization of \citealt{porth19}), the field can push aside infalling plasma and escape.

The unstable equilibrium with $\phi \sim \phi_c$ is known as a magnetically arrested disk \citep[MAD, see][]{BisnovatyiKogan1974, Igumenschchev2003, Narayan2003}, in contrast to accretion flows with $\phi \ll \phi_c$, which are said to follow standard and normal evolution \citep[SANE, see][]{Narayan2012, Sadowski2013}.  Notice that $\phi$ is determined by the nonlinear evolution of the flow and is not trivially related to the initial conditions, although initial conditions have been identified that lead to SANE or MAD outcomes over finite integration times.  We will consider both SANE and MAD accretion flows.

\subsection{Shear at the Jet-Disk Boundary}
\label{sec:jet-disk-shear}

Changes in velocity across the jet--disk boundary may drive Kelvin-Helmholtz instability.  What is the expected velocity difference?  The jet and disk are unsteady and strongly nonaxisymmetric in the numerical GRMHD models that motivate this calculation. In the interest of producing a model that can be studied analytically, we nevertheless treat the system as axisymmetric and steady, and because this is already a drastic approximation, we use a nonrelativistic fluid model for simplicity.

The jet can be idealized as a steady flow anchored in an object rotating with angular velocity $\Omega_F$.  For a steady, axisymmetric, nonrelativistic MHD wind  with plasma angular velocity $\Omega$ and generalized specific angular momentum $L$, angular velocity changes with cylindrical radius $R$ like
\begin{equation}\label{eq:mhdwind}
    \Omega = \Omega_F \dfrac{1}{1 + M_\rA^2} + \dfrac{L}{R^2}\dfrac{M_\rA^2}{1 + M_\rA^2}
\end{equation}
\citep[e.g.,][]{ogilvie2016} where $M_\rA^2 \equiv v_p^2 / v_\rA^2$ is the \alf Mach number,  defined as the ratio of the poloidal plasma velocity to the \alf velocity $\bv_\rA = {\bf B}/\sqrt{\rho}$.  Since $\sigma \gg 1$, $v_\rA \simeq c$.   

Particles flow inward at the horizon and outward at large radius, and therefore a steady state can be achieved only if plasma is loaded onto field lines at intermediate radius.  We assume this occurs, perhaps through turbulent diffusion or through pair production (in numerical GRMHD models plasma is added via numerical floors; see \citealt{wong2021drizzle} for a study of drizzle pair production in this region), and that there is a stagnation point at $r \sim {\rm few} \times G M/c^2$ between an inner, inflow \alf point ($M_\rA^2 = 1$) and an outer, outflow \alf point.  The outer \alf point is close to the light cylinder $r_l \sin\theta = c/\Omega_F$.  

Equation \ref{eq:mhdwind} implies that for $M_\rA^2 \ll 1$, $\Omega \sim \Omega_F$, and for $M_\rA^2 \gg 1$ the specific angular momentum of the wind is conserved.  Inside of the light cylinder, in the limit that $\bhspin \ll 1$, rotation is controlled by the rotation frequency of the hole $\Omega_H$, like $\Omega_F \approx \Omega_H / 2 \approx \bhspin / 8$, so 
\begin{align}
    \Omega \approx 
        \begin{dcases}
        \dfrac{\Omega_H}{2} & r < r_l \\
        \dfrac{2 c^2}{\Omega_H} \dfrac{1}{(r \sin\theta)^2} & r > r_l \\
    \end{dcases}
\end{align}
The jet--disk boundary is at $\theta_{\rm JD}$, so the outer light cylinder radius is $r_l = (8/\bhspin) (G M/c^2)/(\sin \theta_{\rm JD}) + \mathcal{O}(\bhspin)$.  Taking $\sin\theta_{\rm JD} \simeq 1/\sqrt{2}$, then $r_l \simeq (11/a) (G M/c^2)$.  

The disk rotates with approximately constant angular velocity $\Omega = s \Omega_K$ on spherical surfaces;  here $\Omega_K = (G M)^{1/2} r^{-3/2}$ is the Keplerian angular velocity and $0 < s < 1$ measures how sub-Keplerian the accretion flow is. Numerical simulations suggest $s \lesssim 1/2$ for MADs \citep[e.g.,][]{Narayan2012} and $\sim 1$ for SANEs.

The toroidal component of the velocity difference across the jet--disk boundary is thus
\begin{equation}\label{eq:veldiff}
\Delta v_\phi \simeq r\sin\theta_{\rm JD} (\Omega_F - s \Omega_K).
\end{equation}
Without a model for flow along the field lines it is not possible to constrain the other components of the velocity difference.   For retrograde accretion with $\bhspin < 0$, the two angular frequencies in Equation~\ref{eq:veldiff} have the same sign and the magnitude of the velocity jump is at least of order the orbital speed.  The velocity difference is approximately $c$ at $r = r_l$.  For prograde accretion with $\bhspin > 0$, the shear vanishes at $r = 4 (s/\bhspin)^{2/3} (G M/c^2)$, and as in the retrograde case, the velocity difference is $\sim c$ at $r = r_l$.

\subsection{Stability of the Jet-Disk Boundary}

The jet--disk boundary is associated with sharp changes in density and magnetic field.  The jet contains a laminar $\sigma > 1$ plasma, analogous to a pulsar wind, that rotates with the black hole.  The disk contains a turbulent $P_{\mathrm{gas}}/B^2 \sim 1$ plasma whose angular momentum need not be related to the spin of the central hole.  The relative orientation of the shear, jet magnetic field, and disk magnetic field may vary as turbulence in the disk produces varying conditions at the boundary. 

Is the jet--disk boundary linearly stable?  If we model the boundary layer as an infinitely thin current-vortex sheet, then we expect to capture the main features of the linear theory; finite thickness $H$ tends to suppress instability for modes with wavelength smaller than or of order $H$ and fastest growth is at wavelength $\sim H$. The current-vortex sheet can be subject to  Kelvin--Helmholtz instability (KHI) as well as the plasmoid instability (\citealt{Loureiro2007}). High resolution axisymmetric models of black hole accretion flows (\citealt{Ripperda2020}; \citealt{Nathanail2020}) see evidence for plasmoid instability at the jet--disk boundary, but we do not, perhaps due to inadequate resolution.  We therefore focus on KHI.  It is well known that magnetic fields weaken the KHI because they resist corrugation of the vortex sheet. Do magnetic fields stabilize the jet--disk boundary? 

A general linear theory of the plane-parallel, {\em relativistic}, ideal current-vortex sheet does not exist.  \cite{Osmanov2008} consider the special case where magnetic field is oriented parallel to the velocity shear and the density, pressure, and field strength are continuous across the sheet.  They do not consider the large density contrast that is an important feature of the jet--disk boundary problem.

The linear theory of the plane-parallel, compressible, {\em nonrelativistic}, ideal current-vortex sheet is better understood.  The general (arbitrary field orientation on either side of the sheet) incompressible case was considered by \cite{Axford1960}; \cite{Shivamoggi1981} considers aligned and transverse fields; \cite{Sen1964} and \cite{Fejer1964} consider a general, arbitrarily oriented field on either side of the sheet. The stability of a finite-width layer has been considered in a well-known analysis by \cite{MiuraPritchett1982}, but an analytic dispersion relation is not available.  Since the general, nonrelativistic problem is relatively tractable we provide a brief discussion and use it to obtain a qualitative understanding of stability of the jet--disk boundary.

Consider a plane-parallel, nonrelativistic, current-vortex sheet.  The flow velocity and magnetic field are constant away from the sheet, which we position at $z = 0$.  Let $i=J$ denote the low density (jet) side and $i=D$ the high density (disk) side. In equilibrium, ${v_\rA}_z$ vanishes and total pressure is continuous across $z = 0$.

Now consider a perturbation of the form $f(z) \exp(i k_x x + i k_y y + i \omega t)$ with $f(z)=\exp(\kappa z)$, where $\kappa$ is in general complex. 
The general dispersion relation is
\begin{gather}\label{eq:nonrelDR}
    \lambda_J m_D+\lambda_D m_J=0 \\ 
    \lambda_i=\rho_i\left[{\left(\omega-\bk \cdot \bv_i\right)}^2-{\left({\bv_\rA}_i\cdot \bk\right)}^2\right] \\ 
    m_i=\sqrt{k^2+\frac{{\left(\omega - \bk \cdot \bv_i\right)}^4}{{{c_s}_i}^2{\left({\bv_\rA}_i\cdot \bk \right)}^2 - {c_s}_i^2{\left(\omega-\bk \cdot \bv_i\right)}^2}} \label{eq:midefn}
\end{gather}
\citep{Sen1964, Fejer1964}.  Here, $c_s \equiv$ sound speed, $c_m^2 \equiv v_\rA^2 + c_s^2$ is the magnetosonic speed, and $\bv$ is the plasma velocity. The exponential factor $\kappa$ can be $m_i$ or $-m_i$ (see Equation~\ref{eq:midefn}) depending on the boundary condition and whether $z>0$ or $z<0$.

The general dispersion relation cannot be solved analytically.  In the case of interest to us, however,  $\rho_J \ll \rho_D$, ${c_s}_D \sim {v_\rA}_D$, and ${c_s}_J \sim {c_s}_D$.  Furthermore, physics provides a hint to the mathematical solution: the field in the jet is stiff (the \alf speed is large due to the low density), motivating us to look for instability in modes with $\bk \cdot {\bv_\rA}_J = 0$.  This is enough to make analytic progress. Taking $\rho_J/\rho_D \sim \epsilon^2 \ll 1$ and assuming that $\bk \cdot {\bv_\rA}_D \sim \epsilon$, we can solve the dispersion relation to lowest order in $\epsilon$. The relevant mode has 
\begin{equation}\label{eq:KH_NR_LOW_DISP}
\omega^2 = (\bk \cdot {\bv_\rA}_D)^2 - \frac{\rho_J}{\rho_D} \, {\left[\bk \cdot (\bv_J - \bv_D)\right]}^2,
\end{equation}
which suggests that the current-vortex sheet is unstable when $\bk \cdot {\bv_\rA}_D$ is sufficiently small, which we have confirmed by numerically solving the full dispersion relation.  

In Equation (\ref{eq:KH_NR_LOW_DISP}) the nonrelativistic current-vortex sheet is unstable for small $\rho_J$. This is precisely the limit where one might worry about relativistic corrections: if $B_J^2/\rho_J > 1$, then the inertia of the jet is dominated by the magnetic field.  In a fully relativistic analysis (Y.~Du et al., in prep.)~the current-vortex sheet has a near-identical dispersion relation in the limit $\rho_J \rightarrow 0$, except that $\rho_J/\rho_D$ in the above dispersion relation is replaced by $B_J^2/\rho_D$. 

Evidently the current-vortex sheet is not generically unstable at large density contrast: a particular configuration of magnetic fields is needed for instability.  The disk contains a turbulent magnetic field that is constantly changing strength and orientation, while the jet has a steadier field.  This suggests a picture in which turbulent mixing driven by the KHI is episodic and occurs when jet and disk magnetic fields are aligned or anti-aligned.  Mixing as a result of nonlinear development of the KHI will then only occur when there exist modes with growth times that are small compared to the correlation time of the turbulent eddies.    

\subsection{Dissipation at the Jet-Disk Boundary}

The jet--disk boundary would appear to be a fertile setting for particle acceleration:  particles that cross the boundary from the disk plasma frame to the jet plasma frame gain energy in a process akin to Fermi acceleration. This has been investigated by, e.g., \citet{Berezhko1981, Jokipii1990, Ostrowski1990} (see \citealt{Rieger2019} for a review), usually in the context of extragalactic radio jets kiloparsecs from the central source. \citet{Sironi2021Reconnection} performed 2D particle-in-cell simulations of the shear layer between a relativistic, magnetically-dominated electron--positron jet and a weakly magnetized ion--electron plasma and showed that the non-linear evolution of Kelvin--Helmholtz instabilities leads to magnetic reconnection, which can in turn drive particle acceleration. The formation of magnetic islands at the jet--disk boundary \citep[see, e.g.,][]{Ripperda2020, Nathanail2020} can also lead to particle acceleration; this process has been extensively investigated in kinetic simulations of current sheets.

To schematically address this question, we adopt a turbulent resistivity model for dissipation in the jet--disk boundary with magnetic diffusivity $\eta \simeq \alpha\, W \, \Delta v$, where $\alpha$ is the inverse of the magnetic Reynolds number, the width of the boundary layer is $W$ $\sim f R$ ($f < 1$; here, $R \equiv$ cylindrical radius), and $\Delta v \sim c$, so that $\eta \simeq \alpha f c R$.  Next, we assume that the boundary is steady, axisymmetric, and follows $R = R_0 (z/z_0)^\beta$, with the jet intersecting the horizon at $(R_0, z_0)$. We assume that the magnetic flux in the jet $\Phi \simeq \pi B R^2$ is approximately independent of $R$ and thus take $B \simeq \Phi (z/z_0)^{-2\beta}/(\pi {R_0}^2)$.  

If the magnetic field in the disk is similar in magnitude to that in the jet but randomly oriented, the dissipation rate per unit volume in the boundary layer is $\Lambda \sim \alpha B^2 (c/(f R))$, and the total dissipated power per unit height $z$ is independent of $f$:
\begin{equation}
    \frac{d P}{dz} = \frac{1}{\pi}\alpha c \frac{\Phi^2}{R_0^3} \left( \frac{z}{z_0} \right)^{-3 \beta} \left( 1 + \beta^2 \frac{R_0^2}{z_0^2} \left(\frac{z}{z_0}\right)^{-2 + 2 \beta} \right)^{1/2}.
\end{equation}
Notice that this scales asymptotically as $z^{-1 - 2 \beta}$ for $\beta \ge 1$, so nearly all dissipation occurs close to the black hole. Integrating over $z$, the dissipated power is
\begin{equation}
    P = \frac{\alpha c^5\Phi^2}{\pi (G M)^2} F(\beta, z_0/R_0),
\end{equation}
where $F$ is a dimensionless function of order unity.  The power differs only by a factor of $\bhspin^2/\alpha$ from the Blandford--Znajek power (e.g., \citealt{Tchekhovskoy2011}).  To sum up: a fraction $\sim \alpha/\bhspin^2$ of the jet power can be dissipated in the jet--disk boundary close to the black hole; this provides additional motivation for a numerical study.

\section{Simulating Black Hole Accretion}
\label{sec:grmhd}

We now study the jet--disk boundary layer using GRMHD simulations.

\begin{figure*}[th!]
    \centering
    \includegraphics[width=.95 \textwidth]{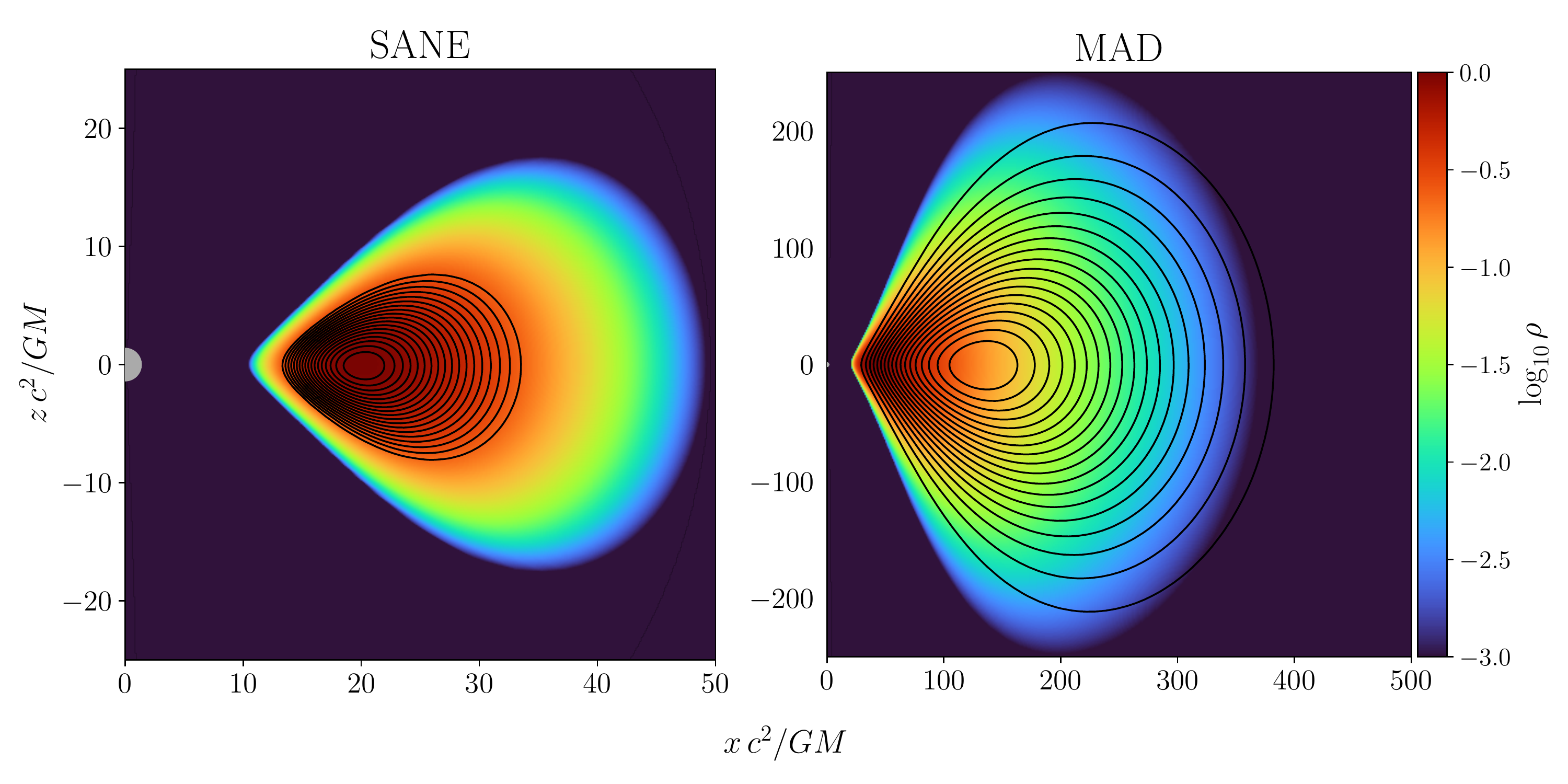}
    \caption{Initial distribution of plasma and magnetic field for representative retrograde SANE (left) and MAD (right) simulations. Both black holes have $\bhspin=-0.94$. The initial plasma density and magnetic field are axisymmetric. The central black hole is plotted at the center left of each panel. Color encodes log$_{10}$ of plasma density, and magnetic field lines, which are purely poloidal, are overplotted in black. Notice that the domain of the MAD plot is $10$x larger than the SANE simulation domain.}
    \label{fig:grmhd_initial_condition}
\end{figure*}

\subsection{Numerical Setup}

We integrate the equations of GRMHD using the \iharm3d code, a descendent of the second order conservative shock capturing scheme {\tt{}harm} \citep{Gammie2003}. Written in a coordinate basis, the governing equations of GRMHD are
\begin{align}
\partial_t \left( \sqrt{-g} \rho_0 u^t \right) &= -\partial_i \left( \sqrt{-g} \rho_0 u^i \right), \label{eqn:massConservation}\\
    \partial_t \left( \sqrt{-g} {T^t}_{\nu} \right) &= - \partial_i \left( \sqrt{-g} {T^i}_{\nu} \right) + \sqrt{-g} {T^{\kappa}}_{\lambda} {\Gamma^{\lambda}}_{\nu\kappa},  \\
\partial_t \left( \sqrt{-g} B^i \right) &= - \partial_j \left[ \sqrt{-g} \left( b^j u^i - b^i u^j \right) \right], \label{eqn:fluxConservation} \\
\partial_i \left( \sqrt{-g} B^i \right) &= 0, \label{eqn:monopoleConstraint} 
\end{align}
where the plasma is defined by its rest mass density $\rho_0$, its four-velocity $u^\mu$, and $b^\mu$ is the magnetic field four-vector following \citet{mckinney04}. Here, $g \equiv {\rm det}(g_{\mu\nu})$ is the determinant of the covariant metric, $\Gamma$ is a Christoffel symbol, and $i$ and $j$ denote spatial coordinates.
In Equations~\ref{eqn:fluxConservation} and~\ref{eqn:monopoleConstraint}, we express components of the electromagnetic field tensor $F^{\mu\nu}$ as $B^i \equiv {^\star\!} F^{it}$ for notational simplicity. The stress--energy tensor ${T^\mu}_\nu$ contains contributions from both the fluid and the electromagnetic field:
\begin{align}
T^{\mu}_{\nu} &= \left( \rho_0 + u + P + b^{\lambda}b_{\lambda}\right)u^{\mu}u_{\nu} \nonumber \\
& \quad + \left(P + \frac{b^{\lambda}b_{\lambda}}{2} \right)g^{\mu}_{\nu} - b^{\mu}b_{\nu},
\end{align}
where $u$ is the internal energy of the fluid and the fluid pressure $P$ is related to its internal energy through an adiabatic index $\hat{\gamma}$ with $P \equiv \left(\hat{\gamma} - 1\right) u$. The \iharm3d code has been extensively tested and converges at second order on smooth flows \citep{Gammie2003}. A comparison of contemporary GRMHD codes can be found in \citet{porth19}.  

Our model has several limitations. First, we treat the accreting plasma as a nonradiating ideal fluid of protons and electrons. We do not consider effects due to anisotropy and conduction (\citet{Sharma2006, Johnson2007}, but see \citet{foucart2017} for an evaluation of the limits of this approximation). We also neglect radiation.  This approximation may be inappropriate in systems with high mass accretion rates, like M87 \citep{Dibi2012, Ryan2017}, but it is sensible in systems with low $\dot{m}$ like Sgr A* (but see \citealt{Yoon2020}, who show a different result under the assumption that the ions and electrons are perfectly coupled). The equations of nonradiative GRMHD are invariant under rescalings of both length and density, so our numerical results can be scaled to the desired $M$ and $\dot{M}$.  

The {\tt{}iharm3d} code evolves plasma on a logically Cartesian grid. For these simulations, we use FMKS coordinates, which are a modified version of the conventional horizon-penetrating Kerr--Schild coordinates. We provide a detailed description of FMKS in Appendix~\ref{sec:fmks}. We use outflow boundary conditions for the radial direction, and we use a reflecting boundary condition at poles that mirrors the elevation components of the magnetic field and fluid velocity across the one-dimensional border. 

We have added a passive tracer particle capability to \iharm3d to track mass loading into the jet.  Each tracer particle is introduced with probability proportional to the coordinate particle density $\sqrt{-g} \rho u^t$, where $\rho$ is the rest-mass density, $g$ is the determinant of the covariant metric, and $u^t$ is the time component of the four-velocity.  Initial positions are uniformly distributed in the coordinate basis in each zone.   Particles are advected with the fluid according to
\begin{align}
    \dd{x^i}{t} = \dfrac{u^i}{u^t},
\end{align}
where $x^i$ are the spatial components of the tracer particle's position and $u^\mu$ is the fluid four velocity. 

The computational cost of evolving the tracer particles alongside the fluid scales linearly with the number of particles; we use $\approx 2^{25}$ particles, and this noticeably increases simulation cost. We therefore use completed GRMHD simulations to identify an epoch of interest, restart the fluid simulation at the beginning of the epoch, initialize the particles, and re-evolve the fluid to the end of the epoch.

The \iharm3d code has several limitations.  It is not robust when $\sigma \gg 1$ \citep[e.g., in the strong cylindrical explosion test in][]{komissarov1999} or when the ratio of the gas pressure to the magnetic pressure $\beta \equiv 2 P_{\mathrm{gas}} / B^2 \ll 1$.  Numerical stability is ensured by imposing artificial ceilings on $\sigma$ and $1/\beta$ in each zone at each timestep, which are enforced by resetting the density or internal energy density to a floor value that depends on position but not on time.  This has a minimal effect on the flow (as can be checked by varying the ceilings), but it does inject particles in the nearly-evacuated funnel region, where $\sigma$ is large and $\beta$ is small.  

\begin{figure}[th]
    \centering
    \includegraphics[width=\linewidth]{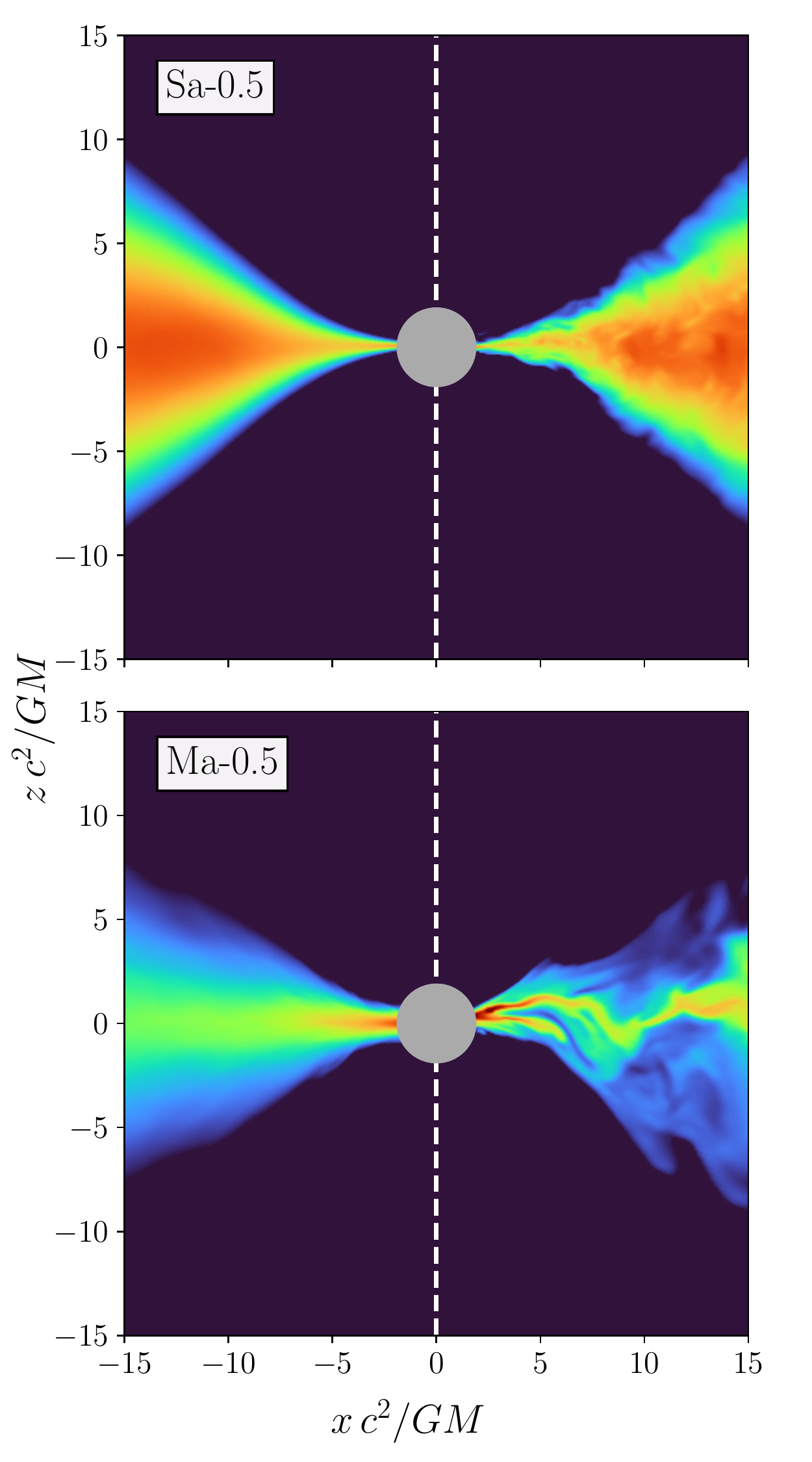}
    \caption{
    Logarithmic plots over three decades of density in the poloidal plane for $\bhspin=-0.5$ MAD and SANE models. Each image shows time- and azimuth- averaged density (left panels) and timeslices at azimuth $\phi = 0$ (right panels). The density is particularly variable in the MAD models, where the timeslice is not well approximated by the average state. The density is less variable in the SANE models, where the timeslice and average state are comparatively similar. 
    }
    \label{fig:azimuthal_snapshot_model_compare}
\end{figure}

\begin{figure*}[th!]
    \centering
    \includegraphics[width=.95 \textwidth]{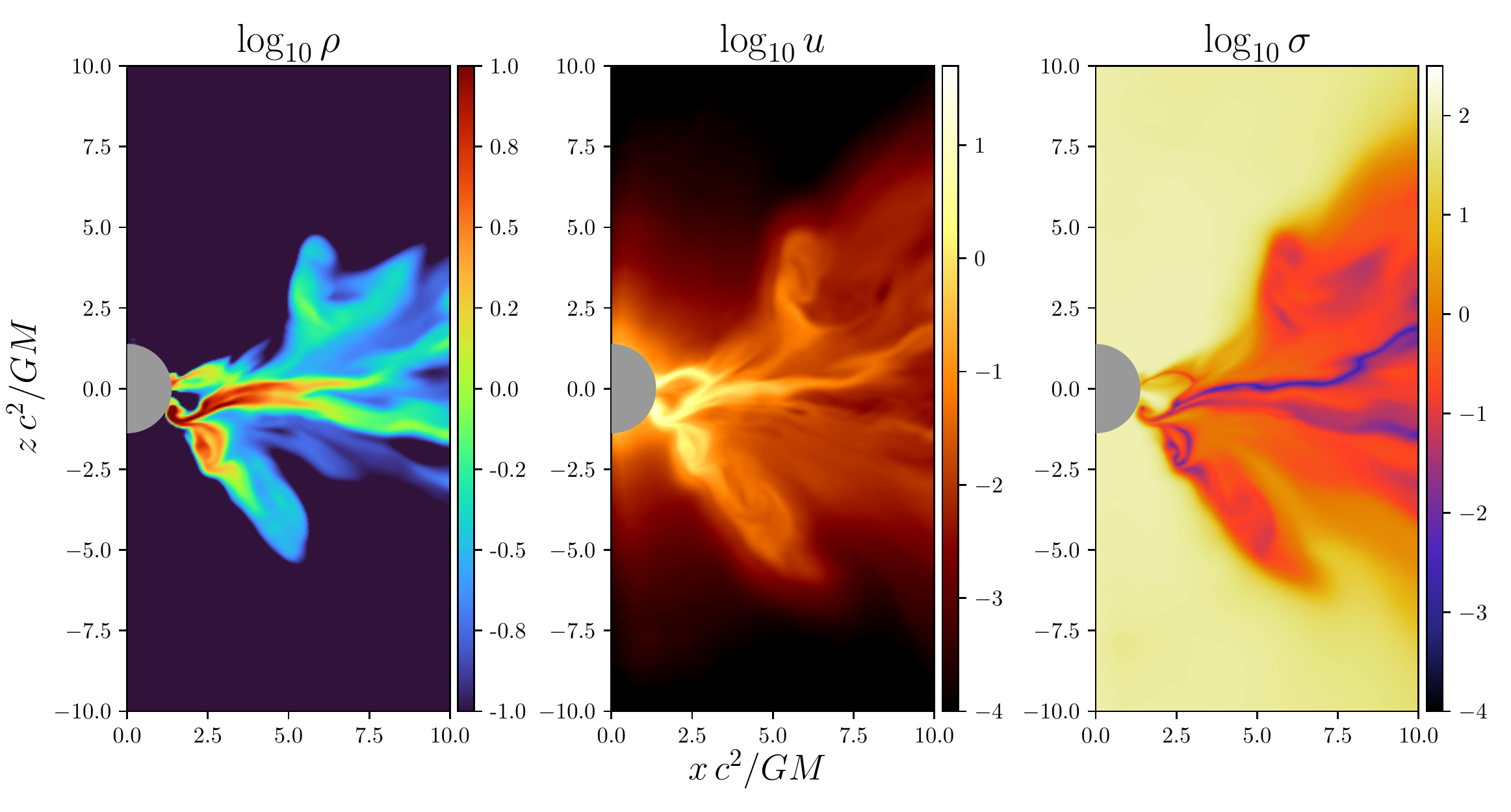}
    \caption{Azimuthal slice from an individual timeslice of the $\bhspin = 0.94$ retrograde MAD simulation. Left panel: log density of plasma near the black hole. Center panel: log internal energy of the plasma $u = \rho T$. Right panel: plasma magnetization $\sigma = b^2/\rho$. The high $\sigma$, low density conical regions around the poles are the jet funnel. The disk is the low $\sigma$, high density region near the midplane. The intermediate region between the funnel and the disk and with $\sigma \approx 1$ is the corona. The disordered accretion near the horizon is accentuated by streams of infalling plasma that are characteristic of MAD accretion.
     } 
    \label{fig:azimuthal_snapshot_cartoon}
\end{figure*}

The fluid sector is initialized with a perturbed Fishbone--Moncrief torus solution \citep{Fishbone1976}, which is parametrized by the inner disk edge radius $r_{\mathrm{in}}$ and pressure maximum radius $r_{\mathrm{max}}$. The thermal energy is perturbed to seed the instabilities that jump start accretion (including the magnetorotational instability). The SANE models have  $r_{\mathrm{in}}=6$ and $r_{\mathrm{max}}=12$ in a domain that extends from within the horizon to $r_{\mathrm{out}}=50 M$. The MAD models have $r_\mathrm{in}=20 M$ and $r_\mathrm{max}=41 M$ in a domain that extends to $r_\mathrm{out}=1000 M$. Our MAD disks are larger than our SANE disks. Figure~\ref{fig:grmhd_initial_condition} shows the initial conditions for plasma and magnetic field in representative SANE and MAD simulations.

The initial magnetic field is described by the toroidal component of the vector four-potential $A_\phi(r, \theta)$. For SANE disks
\begin{align}
    A_\phi = \mathrm{max}\left[\dfrac{\rho}{\rho_{\mathrm{max}}} - 0.2, 0\right],
\end{align} 
where $\rho_{\mathrm{max}}$ is the maximum initial plasma density.
For MAD disks the initial field is concentrated towards the inner edge of the disk and forced to taper at large $r$ according to
\begin{align}
    A_{\phi} = \mathrm{max}\left[\dfrac{\rho}{\rho_{\mathrm{max}}} \left( \dfrac{r}{r_0} \sin \theta \right)^3 e^{-r/400} - 0.2, 0 \right],
\end{align}
where $r_0$ is chosen to be the inner boundary of the simulation domain~\citetext{B.~R.~Ryan, priv.~comm.}.

\subsection{Simulations}

Table~\ref{table:grmhd_models} provides a summary of the models we consider. Our simulations are similar to the retrograde ones generated for the EHT simulation library in \citetalias{PaperV}, except that: our simulations are evolved twice as long to mitigate natural stochasticity in matter entrainment; and a subset of our simulations are rerun at multiple resolutions. 

We focus on four retrograde simulations with $\bhspin = -0.5$ or $-0.94$. By convention, negative spins means that the black hole spin is anti-parallel to the angular momentum of the accretion flow (i.e., tilt is $180$deg). For each spin, we consider MAD and SANE models.  We set the magnetic flux (and thus MAD or SANE state) by varying the field structure in the initial conditions.  

Each simulation was run for at least 20,000 $GM/c^3$ and has an initial transient phase during which the initial torus relaxes, and magnetic winding and a combination of Rayleigh--Taylor and Kelvin--Helmholtz instabilities operate. The transient phase is followed at each radius by a turbulent quasi-equilibrium, with equilibrium radius, defined as the largest radius where $d\dot{M}/dr \simeq 0$, increasing as $r_{\mathrm{eq}} \sim t^{2/3}$ (see, e.g., \citealt{penna2010,dexter2020inflow} for a discussion).  Beyond $r_{\mathrm{eq}}$, the flow is strongly dependent on initial conditions, so we consider information only from $r < r_{\mathrm{eq}}$.  GRMHD models may be in equilibrium at large radii near the poles if there are strong outflows {\em and} the outflow structure is independent of the structure of the surrounding unequilibrated disk.

Our MAD simulations are run with bulk fluid adiabatic index $\Gamma = 13/9$, and our SANE simulations are run with $\Gamma = 4/3$ to be in agreement with \citetalias{PaperV} and \citet{porth19}.

\begin{deluxetable*}{ lllllllc }
\tablecaption{GRMHD Simulation Parameters} \label{table:grmhd_models}
\tablehead{ 
\colhead{id} &
\colhead{flux} & 
\colhead{$\bhspin$} &
\colhead{$r_{\mathrm{in}}$} &
\colhead{$r_{\mathrm{max}}$} &
\colhead{$r_{\mathrm{out}}$} &
\colhead{resolution} &
\colhead{notes} 
}
\startdata
Sa-0.5 & SANE & $-0.5$ & $6$ & $12$ & $50$ & 288x128x128 & medium disk \\
Sa-0.94 & SANE & $-0.94$ & $6$ & $12$ & $50$ & 288x128x128 & medium disk \\
Ma-0.5 & MAD & $-0.5$ & 20 & 41 & $1000$ & 384x192x192 & large disk \\
Ma-0.94\_192 & MAD & $-0.94$ & 20 & 41 & $1000$ & 192x96x96 & large disk \\
Ma-0.94\_288 & MAD & $-0.94$ & 20 & 41 & $1000$ & 288x128x128 & large disk \\
Ma-0.94$^\dagger$ & MAD & $-0.94$ & 20 & 41 & $1000$ & 384x192x192 & large disk, multiple realizations, tracer particles \\
Ma-0.94\_448 & MAD & $-0.94$ & 20 & 41 & $1000$ & 448x224x224 & large disk \\
\enddata
\tablecomments{Retrograde GRMHD fluid simulations parameters. Flux labels the relative strength of the magnetic flux at the horizon, $\bhspin$ describes the spin of the black hole, $r_{\mathrm{in}}$ and $r_{\mathrm{max}}$ are parameters for the initial Fishbone--Moncrief torus, $r_{\mathrm{out}}$ is the outer edge of the simulation domain, resolution gives the $N_r \times N_\theta \times N_\phi$ number of grid zones in the simulation.
$^\dagger$ The 384x192x192 MAD $\bhspin=-0.94$ simulation was run using a different perturbed initial condition, and passive tracer particles were tracked for a part of its evolution.}
\end{deluxetable*}

\section{Results}
\label{sec:results}

We begin by discussing characteristic differences between MAD and SANE accretion flows before considering each of our simulations in detail.  We explore the properties of fluid flow at small radii and within the jet, and then we relate outbursts in the MAD flows to magnetic flux ejection events. We explore qualitative features of the jet--disk boundary layer, including the development of Kelvin--Helmholtz instability. Finally, we use tracer particles to study mass entrainment across the jet--disk boundary layer.

\subsection{Overview}

It is convenient to divide low-luminosity black hole accretion flows into three regions: (1) the matter-dominated disk of plasma near the midplane, which on average flows inward, (2) the magnetically dominated, polar Poynting jet, and (3) the virial temperature intermediate region that contains the jet--disk boundary layer and the corona (here defined as the region with $\beta \sim 1$). 
In a region extending from the event horizon out to somewhat beyond the innermost stable circular orbit (ISCO), the inflow plunges supersonically onto the hole and fluctuates strongly. Notice that the jet we consider here (at horizon scales) is dynamically distinct from the jet at large radius. 

\begin{figure}[th]
    \centering
    \includegraphics[width=\linewidth]{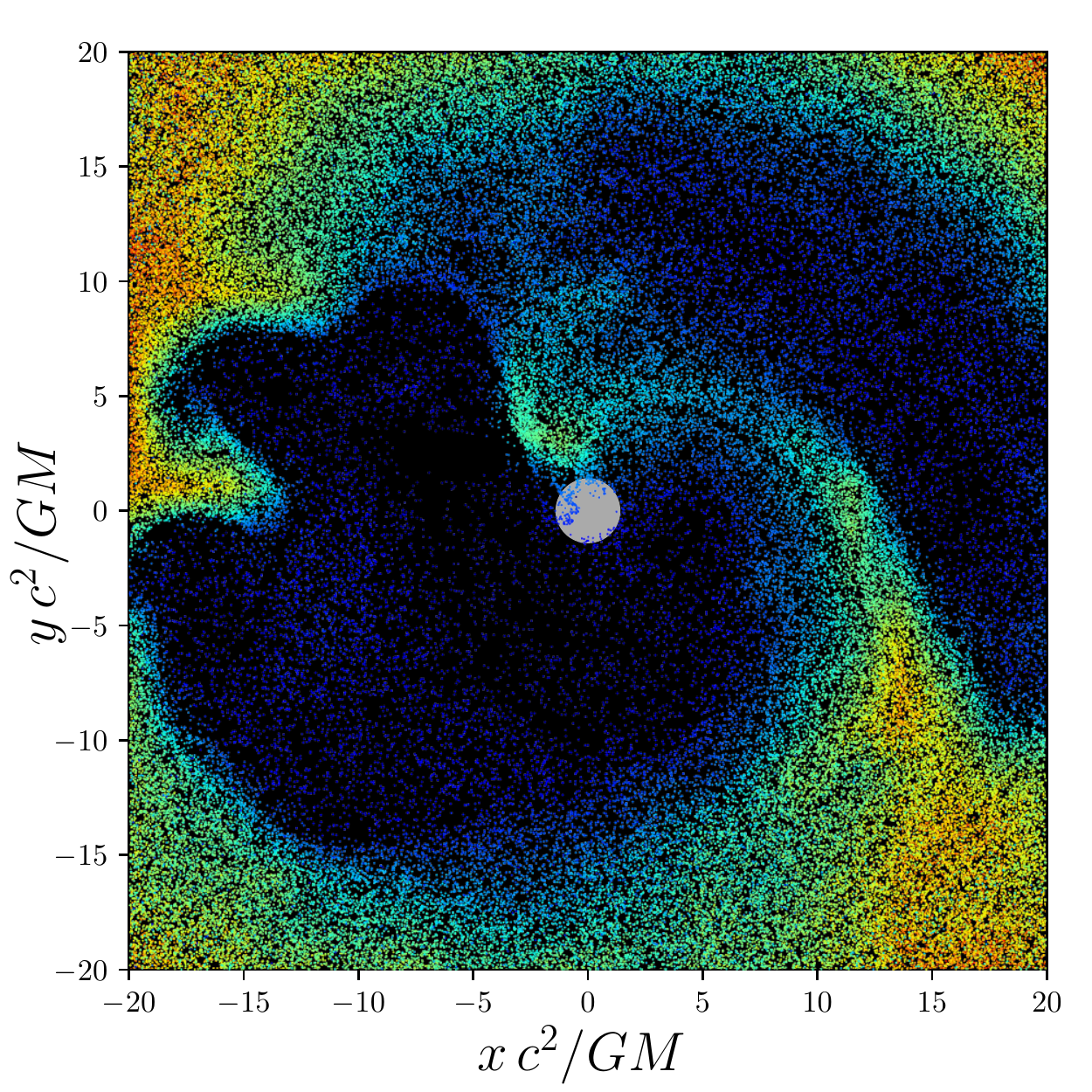}
    \caption{
    Tracer particle position in the MAD, $\bhspin=-0.94$ model, projected onto the equatorial plane. Particle color varies linearly with local rest-mass density. The event horizon is a gray sphere. The inner region of the accretion flow is chaotic and characterized by plasma streams that break off the main disk at large radius.  Plasma streams experience large magnetic torques ($u_\phi$ may change sign) as they plunge toward the horizon.
    }
    \label{fig:grmhd_tracer_fallingin}
\end{figure}

\begin{figure*}[th]
    \centering
    \includegraphics[width=.48\textwidth]{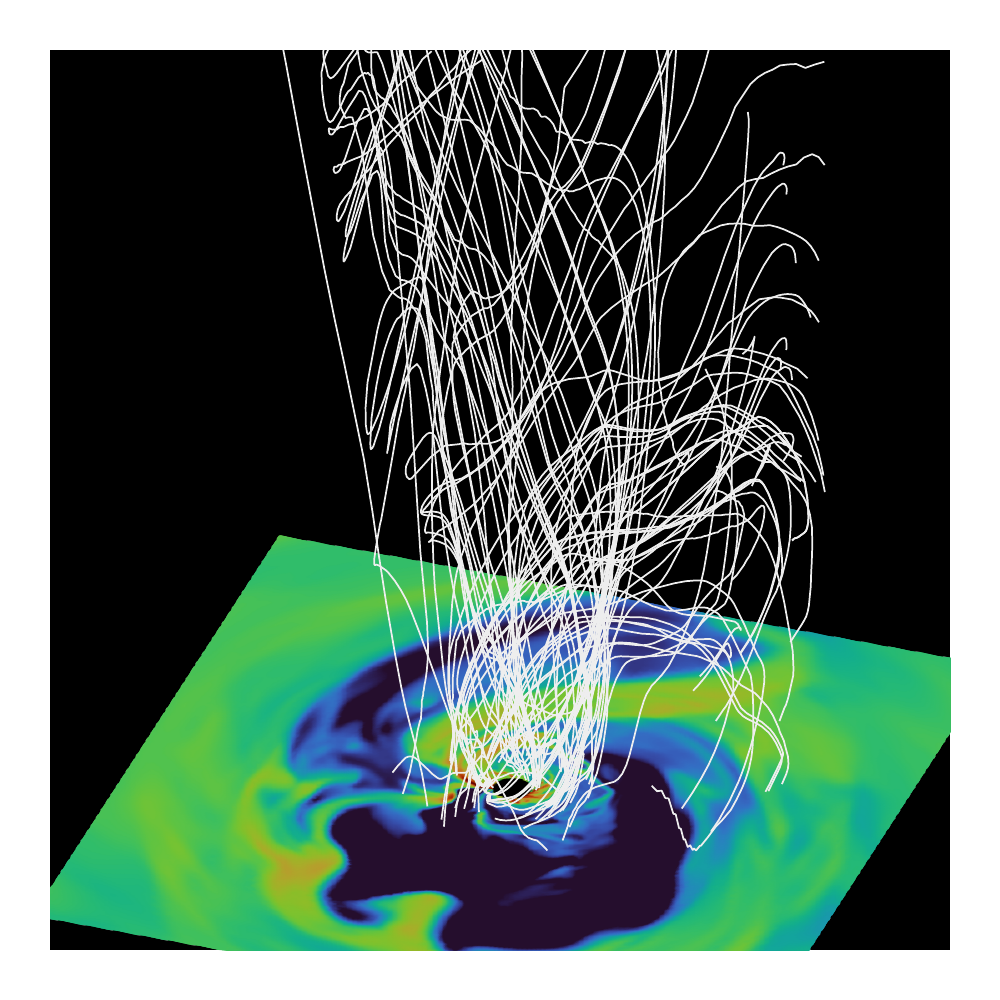}
    \includegraphics[width=.48\textwidth]{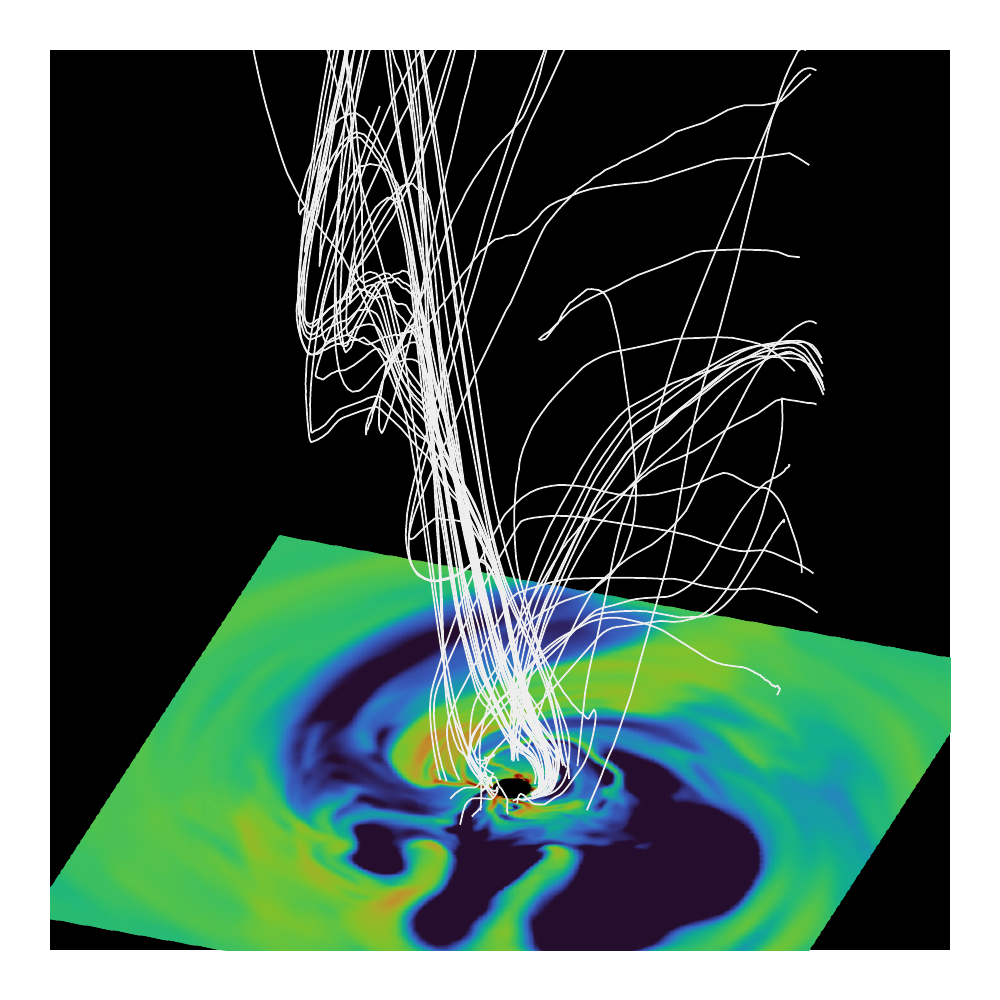}
    \caption{
    Interaction between disk and jet magnetic field lines. 
    Magnetic field lines that intersect the disk at small radii are shown for two sequential timeslices of the plasma evolution. Field lines are sampled according to magnetization in the midplane. The colored surface shows the logarithm over two decades of density in the midplane of the simulation, and the event horizon is plotted as a black circle in the center of the plane. Left panel: the same timeslice as shown in Figure~\ref{fig:grmhd_tracer_fallingin}, rotated $45^\circ$ counter-clockwise. Magnetic field lines emanating from the high density region towards the left of the figure trace an accretion stream and are disk-dominated. Magnetic field lines that wind the opposite direction make up a flux tube and are being pulled clockwise with the hole as it spins. The two sets of field lines are about to collide.
    Right panel: same simulation approximately $50\, GM/c^3$ later. Disk-threading and funnel-threading magnetic field lines have interacted, and a much stronger flux tube passes through the midplane in the low density region to the right of the hole. 
    }
    \label{fig:grmhd_flux_tubes}
\end{figure*}

SANE and MAD accretion flows exhibit qualitatively different behavior. SANE models are relatively tame: plasma falls uniformly from the ISCO to the event horizon, the boundary of the accretion disk remains well defined, and the time-averaged accretion state is a fair approximation of an individual timeslice. In contrast, MAD accretion is choppy and tends to proceed in isolated, thin plasma streams that begin far from the hole and plunge onto it. MAD accretion is punctuated by violent eruptions that release excess trapped magnetic flux. Although the flux ejection events are not understood in detail, their structure suggests a Rayleigh--Taylor interaction between the disk and hole \citep[see, e.g.,][]{Marshall2018}. 
For MAD flows, the time average is often not a good approximation to a single timeslice. These differences are particularly apparent in Figure~\ref{fig:azimuthal_snapshot_model_compare}, which shows log density for sample SANE and MAD models and compares the time-averaged solution (left) to representative timeslices (right).  
In SANE models it is easy to separate the high-density disk from the low-density jet region.  In contrast, in MAD models, identifying the location of the jet--disk boundary is a challenge.

In Figure~\ref{fig:azimuthal_snapshot_cartoon}, we show a typical timeslice on a poloidal slice of an $\bhspin = -0.94$ MAD model, where the strength of the magnetic flux near the horizon prevents steady disk-accretion. 
Here, accretion occurs when plasma streams break from the bulk disk at large radius and plunge onto the hole.
These streams are not confined to the midplane as they fall. Figure~\ref{fig:grmhd_tracer_fallingin} shows the projected locations of tracer particles in the same MAD $\bhspin=-0.9373$ flow of Figure~\ref{fig:azimuthal_snapshot_cartoon} but viewed from above. The color of each particle corresponds to the linear density of particles in a three-dimensional voxel of space centered at the particle and is used to visualize the complicated vertical structure of the flow.
The figure shows one accretion stream connecting the disk and the hole in the bottom right and the launch of two new streams in the upper right.

\begin{figure}[t]
    \centering
    \includegraphics[width=\linewidth]{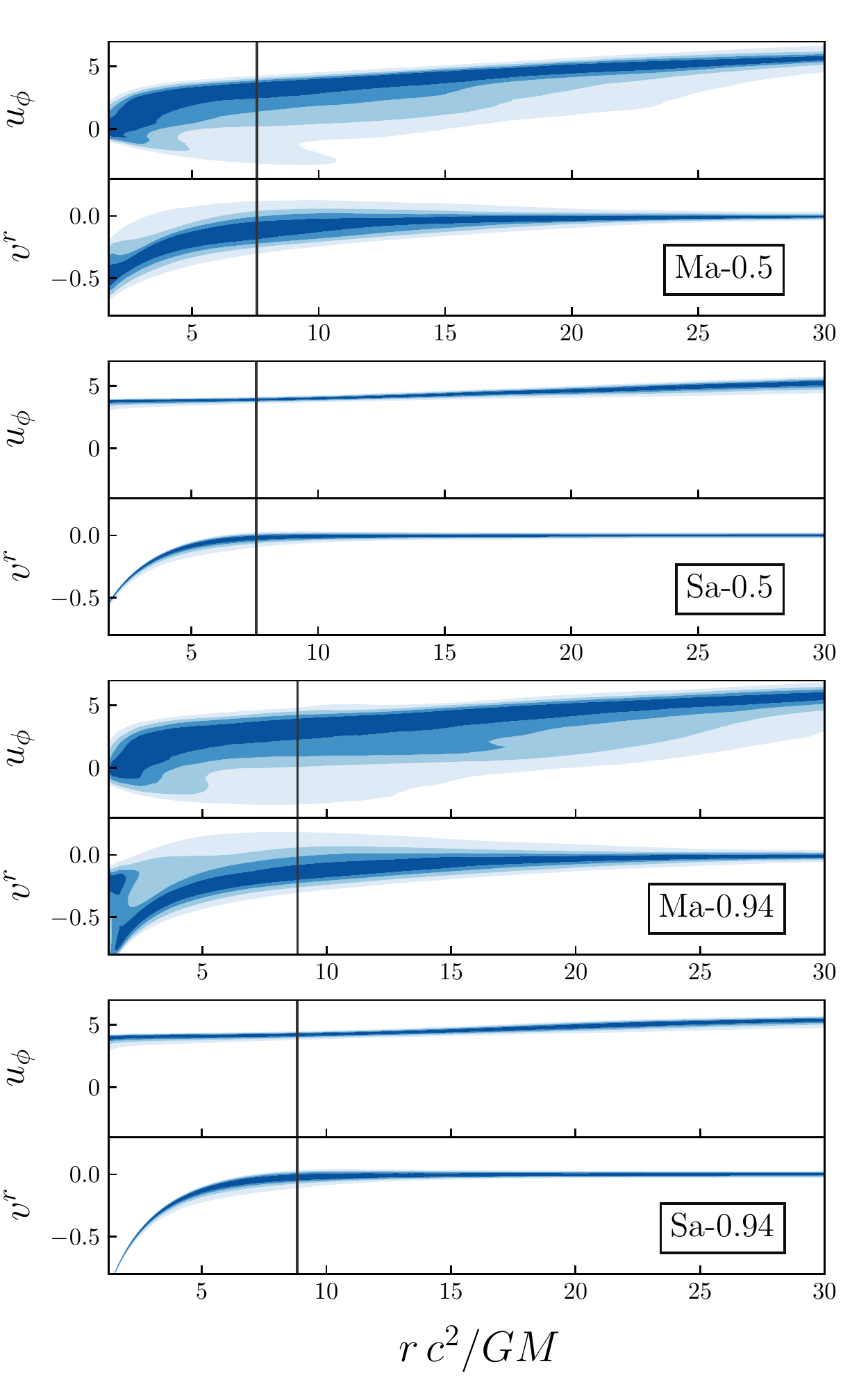}
    \caption{
    Distribution of matter in the angular momentum and radial velocity versus radius ( $u_\phi-r$ and $v^r-r$) planes for the four fiducial simulations. The vertical gray line marks the ISCO. The colorscale is linear and shows the distribution of matter at each radius. In the SANE models the plasma lies on a well defined curve associated with Keplerian rotation as it accretes.  In the MAD models plasma is perturbed away from the disk even before it enters the plunging region.
    }
    \label{fig:grmhd_statehistogram}
\end{figure}

\begin{figure}[t]
    \centering
    \includegraphics[width= \linewidth]{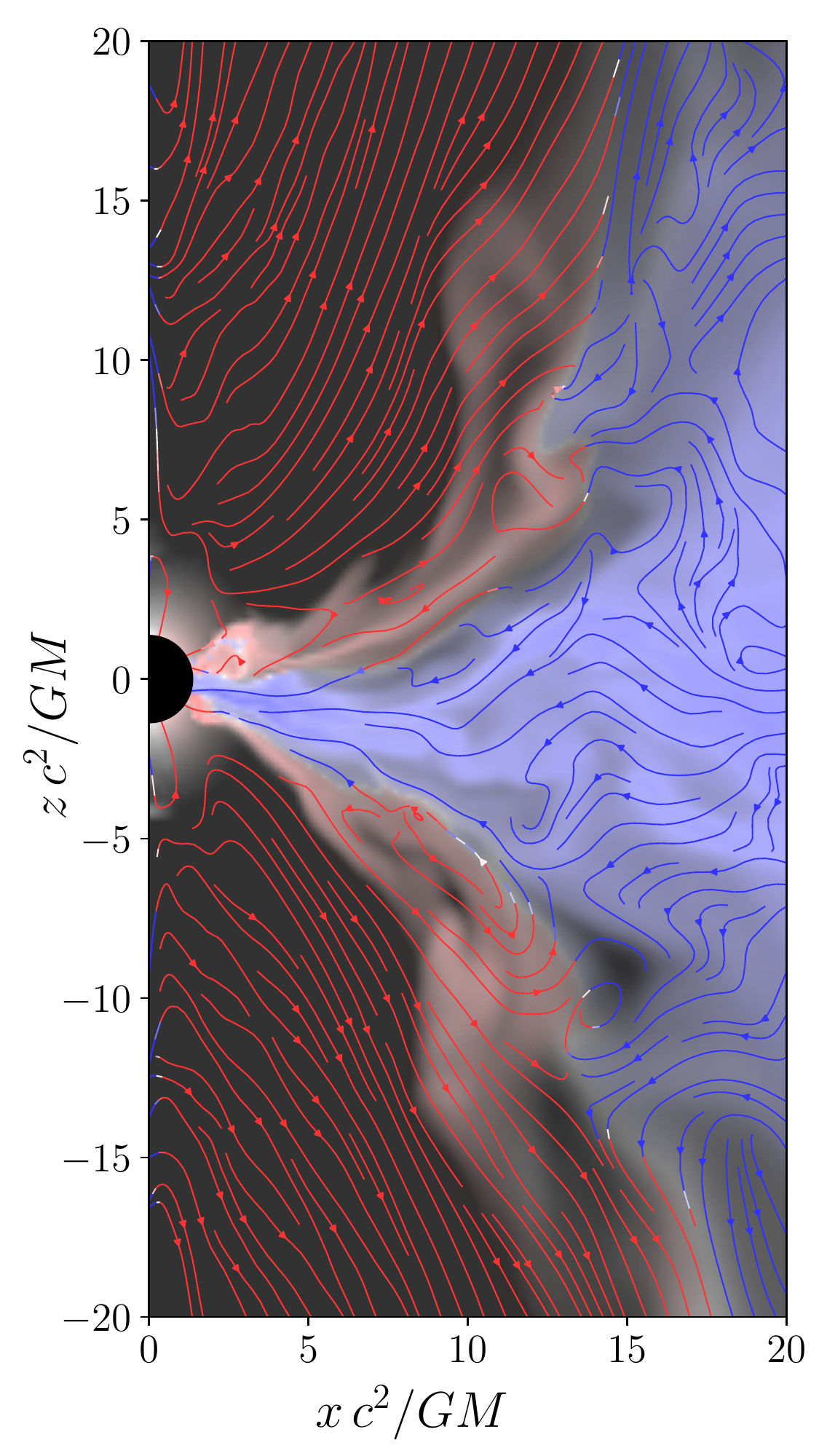} 
    \caption{Timeslice of a MAD, $\bhspin = -0.94$ model. Brightness shows plasma density, color saturation encodes value of $u_\phi$, and flow lines describe the poloidal motion of the plasma. The jet--disk boundary is visible as the surface where $u_\phi$ changes sign.  Eddies tend to form at the jet--disk boundary as infalling, positive $u_\phi$ matter interacts with outflowing, negative $u_\phi$ matter. The sign of $u_\phi$ in the funnel is set by the sign of black hole spin.
    }
    \label{fig:grmhd_shear_layer}
\end{figure}

\begin{figure}[t]
    \centering
    \includegraphics[width= \linewidth]{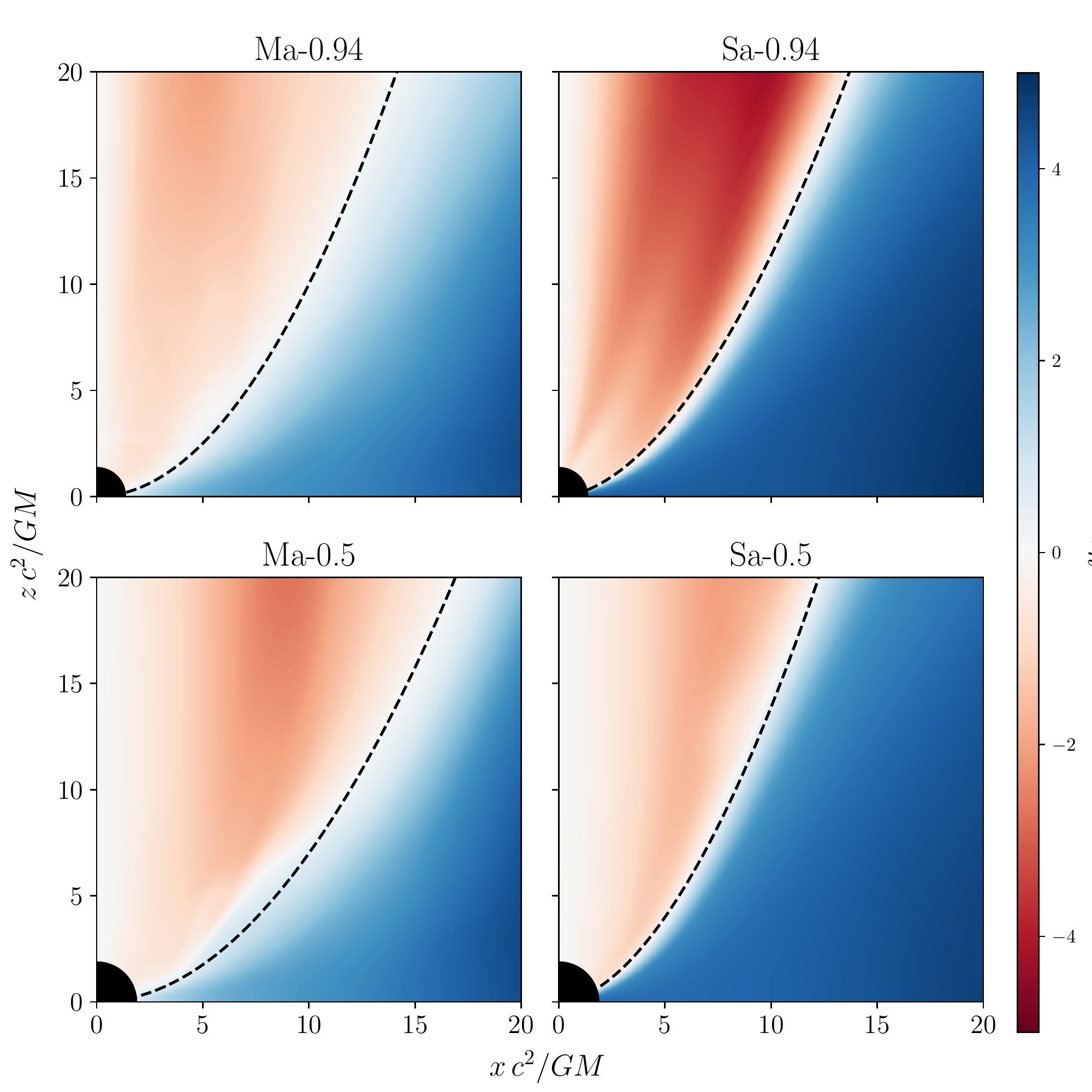} 
    \caption{
    Density--weighted poloidal profile of $u_\phi$ for each of the four fiducial models after time and azimuthal averaging. The black circle at the origin marks the extent of the event horizon. All simulations have a similar structure: a parabolic jet (boundary defined by $u_\phi = 0$) and a peak in $u_\phi$ away from the pole.  
    }
    \label{fig:grmhd_funnel_profiles_2d}
\end{figure}

\begin{figure}[t]
    \centering
    \includegraphics[width= \linewidth]{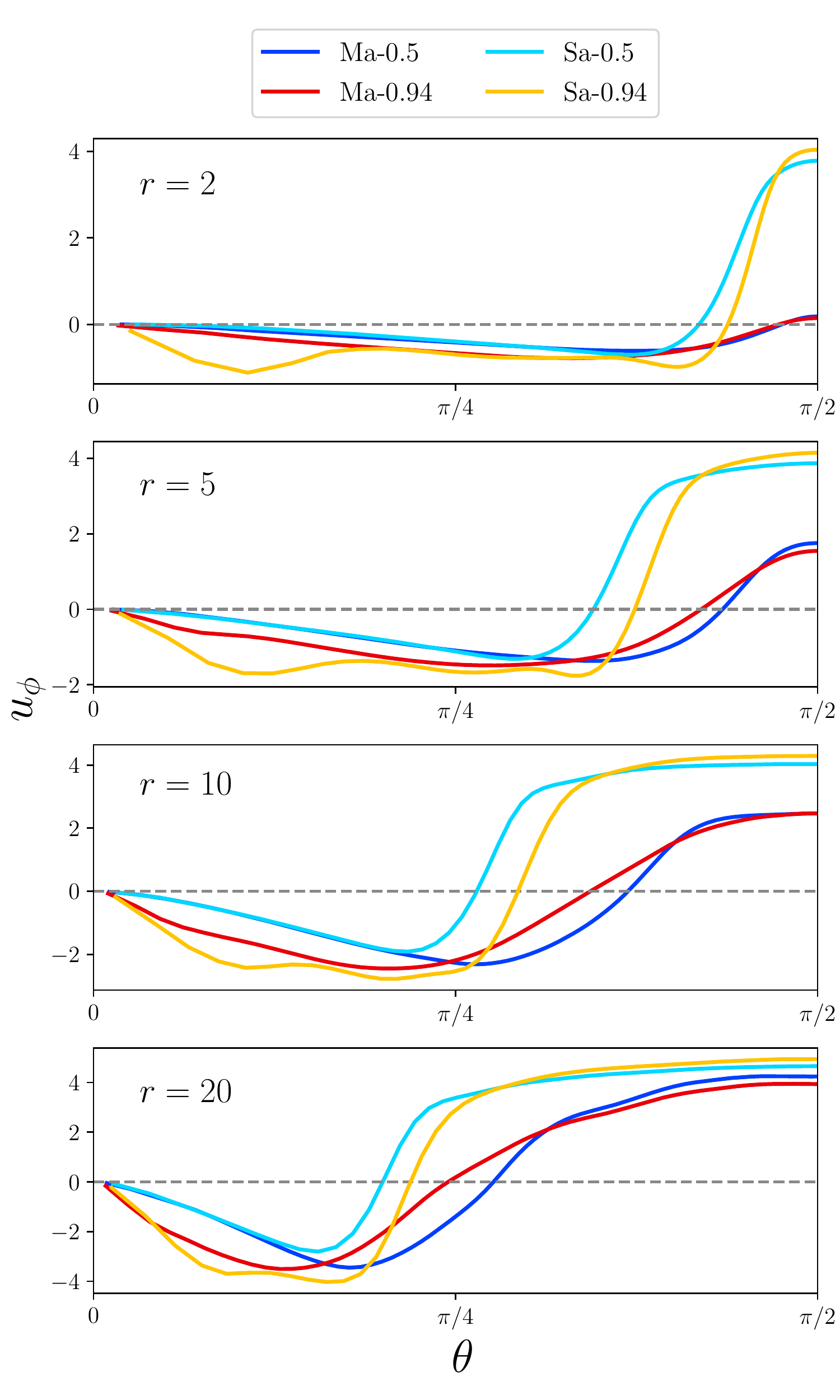} 
    \caption{
    Profile of $u_\phi$ versus elevation at $r=2, 5, 10,$ and $20\;GM/c^2$ for each of the models in Figure~\ref{fig:grmhd_funnel_profiles_2d}.  Notice that $u_\phi < 0$ implies angular momentum aligned with the black hole. The average $u_\phi$ of plasma at small radii is smaller in MAD models than SANE models. The latitude of the shear layer within which $u_\phi$ changes sign increases with radius, corresponding to a narrowing jet. The (average) shear layer is wider for MAD models because their  jet--disk boundary fluctuates over a wider range in latitude. As matter flows out in the jet, magnetic torques increase $u_\phi$.
    }
    \label{fig:grmhd_funnel_profiles}
\end{figure}

\subsection{Counterrotation and the disk}

As the black hole rotates, trapped magnetic field lines wind around the polar axis and produce a Poynting jet via the BZ mechanism.
In the jet--disk boundary layer, however, the jet field lines (that rotate with the hole) are mixed with disk field lines (that rotate against the hole in retrograde models).  This interaction leads to an exchange of angular momentum via magnetic and fluid stresses.  Some of the infalling plasma then acquires negative $u_\phi$, i.e., its specific angular momentum aligns with the black hole spin. 

Exchange of angular momentum in the jet--disk boundary layer is more noticeable in MAD models, where accretion occurs in streams and where the magnetic field tends to be stronger. In MAD models, the inhomogeneous flow magnifies the effects of magnetic torques, since some equator-crossing field lines are lightly loaded (in contrast to SANE models, in which the equator-crossing field lines pass through a dense disk).  Moreover, the more concentrated magnetic flux tubes in the MAD models can result in stronger torques \citep[see][]{porth2020}: when matter in the accretion stream with $u_\phi > 0$ interacts with a flux tube with $u_\phi < 0$, the plasma is rapidly braked and its angular momentum is reversed. Figure~\ref{fig:grmhd_flux_tubes} shows an example of this interaction as counterrotating field lines collide with the corotating field lines near the horizon. During these events, the front edge of an accretion stream commonly erodes and accelerates radially outwards. 

The stronger angular momentum transfer in MAD flows produces more disorder in the inner region of the accretion flows. 
This difference between MAD and SANE models can be seen in Figure~\ref{fig:grmhd_statehistogram}, which plots the time-integrated distributions of rest-mass over $u_\phi, r$ and $v^r, r$. The infalling matter accelerates within the plunging region (close to the ISCO) in both MAD and SANE flows, but the widths of the distributions of $u_\phi$ and $v_r$ at a given radius differ sharply: the MAD models have larger width because they experience larger fluctuations.

\subsection{Jet wall shape}
\label{sec:funnel_profiles}

In general, it is challenging to identify the jet--disk boundary since there is no clear criterion that distinguishes matter in the jet from matter in the disk (although proxy surfaces derived from magnetization or the Bernoulli parameter have been used in the past). Nevertheless, it is straightforward to find the surface where $u_\phi = 0$.  Since $u_\phi$ has a definite sign in the jet, this surface may be a reasonable tracer of the boundary. 

Figure~\ref{fig:grmhd_shear_layer} shows an azimuthal timeslice of plasma density and angular momentum in the MAD $\bhspin=-0.94$ simulation and overplots the flow of the plasma. The lines change color at the $u_\phi=0$ surface, which broadly separates outgoing matter from infalling matter. The extended jet--disk boundary is turbulent and mixes mass, angular momentum, and energy between the two regions. Figure~\ref{fig:grmhd_funnel_profiles_2d} plots time and azimuth averaged $u_\phi$ for each of six models.  We fit the $u_\phi = 0$ surface (within $r < 30\, GM/c^2$) to $z = a x^b$ and plot it as a dashed line. Recall that the boundary produced from ($\phi,t$)-averaged data may not be a good approximation to the boundary at fixed $\phi, t$, especially for MAD models. The parameters for the fit are reported in Table~\ref{table:funnel_wall_parameters}.

\begin{table}[t!]
    \caption{Funnel wall ($u_\phi = 0$ surface) fit parameters}
    \begin{center}
    \tabcolsep=0.09cm
    \begin{tabularx}{0.8\columnwidth}{ c c c }
    \hline\hline
    \hspace{.75cm}id &\hspace{.75cm} $a$\hspace{.75cm} & \hspace{.75cm}$b$\hspace{.75cm} \\
    \hline
\hspace{.75cm}Sa-0.5 & $0.22$ & $1.8$ \\
\hspace{.75cm}Sa-0.94 & $0.18$ & $1.8$ \\
\hspace{.75cm}Ma-0.5 & $0.07$ & $2$ \\
\hspace{.75cm}Ma-0.94 & $0.1$ & $2$ \\
    \hline
    \end{tabularx}
    \label{table:funnel_wall_parameters}
    \end{center}
    \tablecomments{Best fit parameters of $z = A x^b$ model for the location of the zero angular momentum surface in the GRMHD models.}
\end{table}

In Figure~\ref{fig:grmhd_funnel_profiles} we plot $\langle u_\phi\rangle$, where the brackets indicate an average over time and azimuth versus elevation at four radii in each of the simulations. 
In MAD flows, we see that the average $u_\phi$ of matter in the midplane at $\theta = \pi/2$ decreases with radius; this makes sense since horizon-scale accretion flow is much choppier in MADs.
The average $u_\phi$ of the plasma tends to increase with radius in both the disk and in the funnel. The point where $u_\phi$ changes sign corresponds to the location of the jet--disk boundary layer and roughly tracks the shape of the jet. In our SANE simulations, the boundary layer is resolved by $\gtrsim 16$ zones at all radii, and the jet spans approximately $10$ zones at $r=20\,GM/c^2$ and approximately $40$ zones at $r=2\,GM/c^2$. The boundary layer in our MAD simulations spans approximately $\gtrsim 30$ zones at all radii, and the jet is resolved by between $20$ and $60$ zones at $r=20\,GM/c^2$ and $r=2\,GM/c^2$ respectively.

\begin{figure*}[t]
    \centering
    \includegraphics[width=.90 \textwidth]{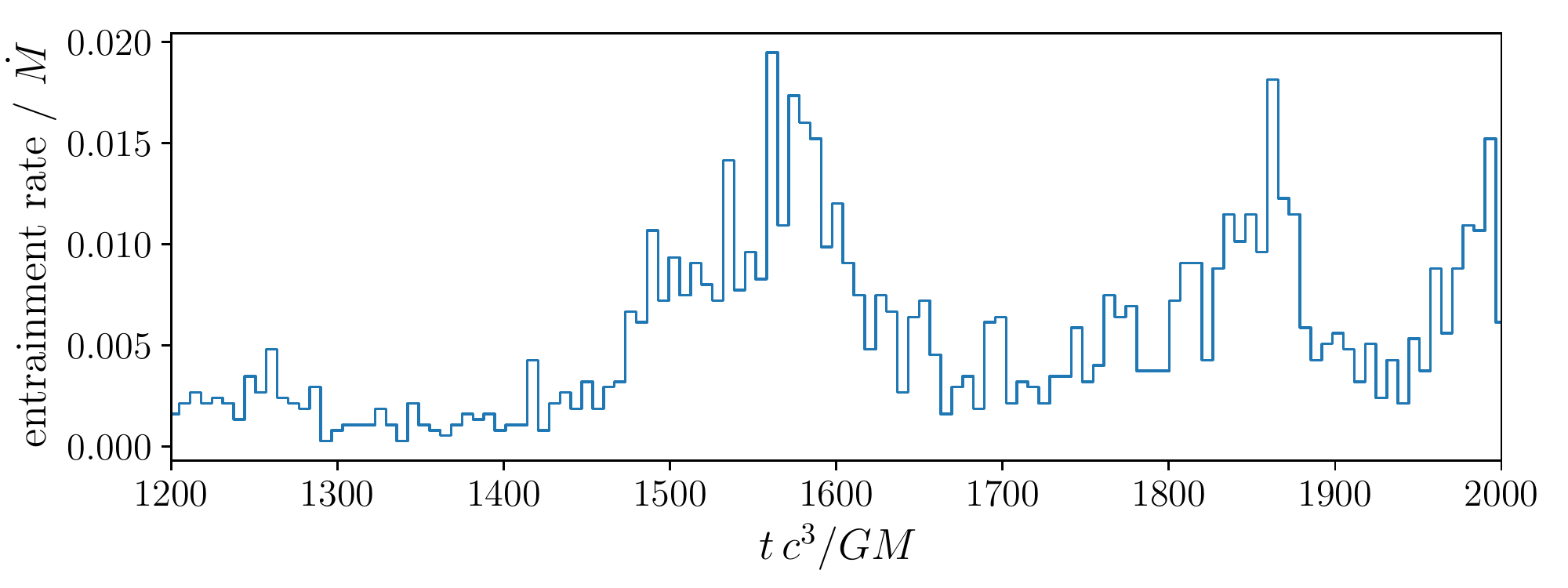}
    \caption{Histogram showing when tracer particles are entrained into the jet over a brief interval in the MAD $\bhspin = -0.94$ model. Entrainment is conservatively defined to only include particles that begin in the disk region and end at large radius with positive $v^r$. This definition discounts particles that spend time in the mixing region but ultimately fall onto the hole. In this MAD model and by these criteria, entrainment is evidently a stochastic process that is characterized by periods of increased entrainment corresponding to times when instabilities form and break at horizon scales.
    }
    \label{fig:grmhd_tracer_entrainment_times}
\end{figure*}

\begin{figure*}[t]
    \centering
    \includegraphics[width=0.98\linewidth]{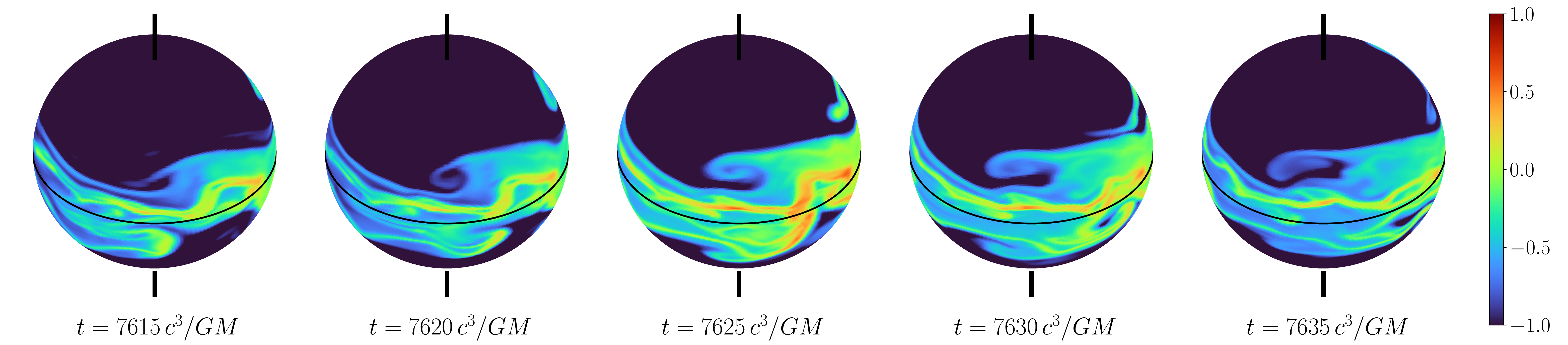}
    \caption{
    Logarithm over two decades of density on $r\approx 1.5 M$ slices for the MAD $\bhspin = -0.94$ model at five times separated by $\Delta t = 25 M$. Matter in the jet near the poles flows clockwise from above (left on the page), and matter in the midplane flows counterclockwise (right on the page). The boundary between the funnel and the midplane results in the development of an unstable shear layer. A Kelvin--Helmholtz roll develops in the shear layer over the sequence of panels.
    }
    \label{fig:grmhd_projection}
\end{figure*}

\begin{figure}[h]
    \centering
    \includegraphics[width=\linewidth]{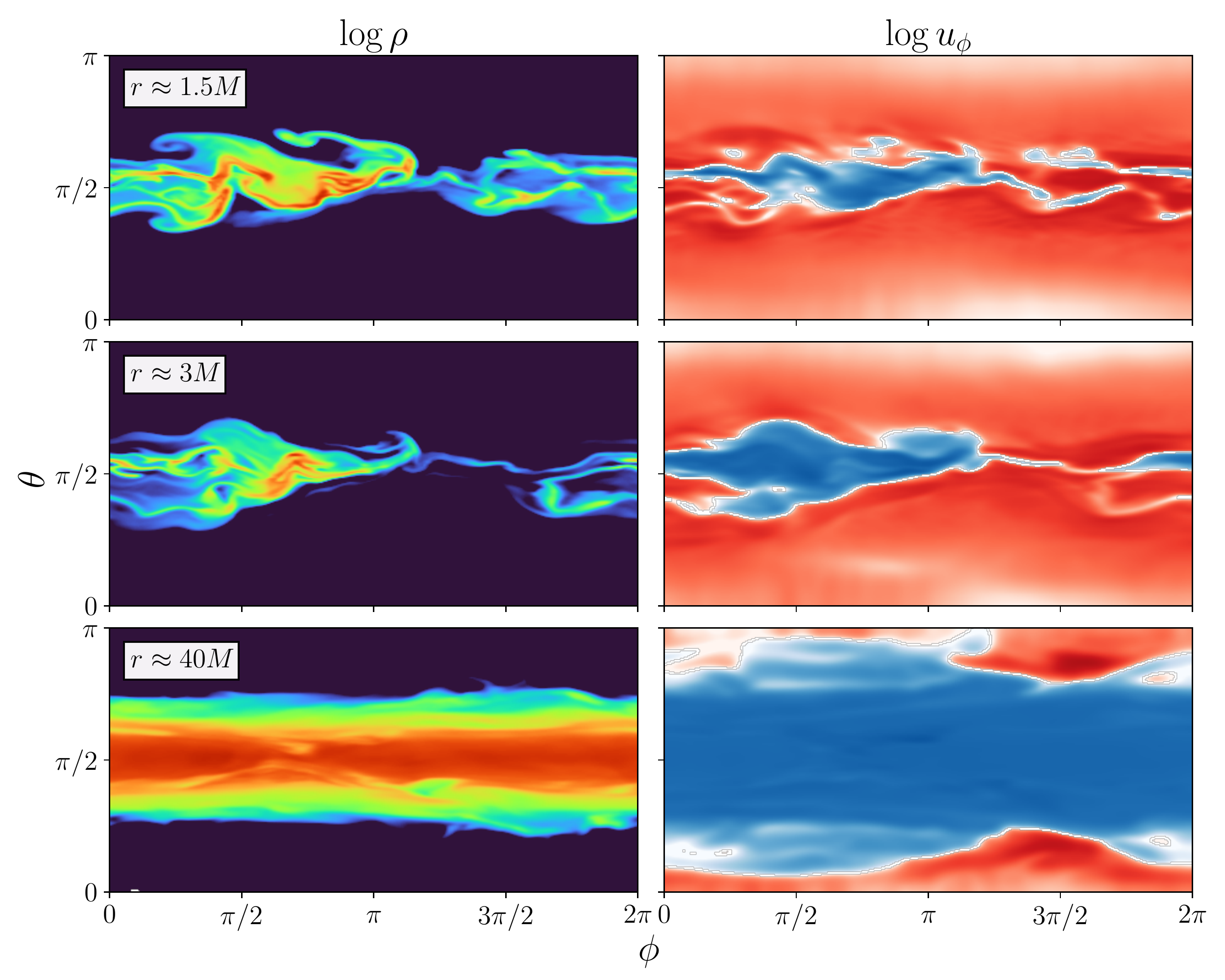}
    \caption{
    Left panels: log over two decades of density in the $\theta-\phi$ plane for shells at $r=1.5, 3, 40\, GM/c^2$.
    Right panels: same shells as left showing logarithm over two decades of $u_\phi$ with $u_\phi >0$ blue and red otherwise. These plots are from the central time slice of Figure~\ref{fig:grmhd_projection}, for the MAD $\bhspin=-0.94$ model.  The flow becomes increasingly chaotic at smaller radii; however, the shear layer between the disk and funnel persists, and the funnel region consistently has $u_\phi < 0$, indicating corotation with the hole.
    }
    \label{fig:grmhd_equirectangular}
\end{figure}

\subsection{Mass entrainment}
\label{sec:mass-entrainment}

The shear layer at the jet--disk boundary is episodically unstable in our models. As instabilities develop, plasma from the disk is transported across the boundary, reverses direction, and is entrained into the jet. We use tracer particles to study mass entrainment and track matter that passes through the mixing region. The computational cost of tracking tracer particles in the global flow over the course of the entire simulation makes a full study prohibitively expensive. We instead perform a single high-resolution, high-cadence study that focuses on the evolution of approximately $3.2\times 10^6$ particles within the inner region of the accretion flow over a $500\, GM/c^3$ interval. We chose to consider a range of time in the MAD $\bhspin=-0.94$ model because it corresponded to an active period when multiple KHI knots are easily identifiable.

Entrained particles satisfy two criteria: they begin with $v^r < 0$ and $u_\phi > 0$, and they leave the simulation at the outer boundary with $u_\phi < 0$. In the mixing layer tracer particles may repeatedly transition between the disk and jet; we define entrainment to have happened for a tracer particle when its $u_\phi$ and $v^r$ change sign for the last time. 
Because this definition of entrainment depends on the worldline of a fluid parcel, it is not immediately analogous to any quantity that can be directly computed from the raw fluid data.

Figure~\ref{fig:grmhd_tracer_entrainment_times} shows the computed mass entrainment rate over time. We find that entrainment events occur in bursts lasting $\sim 100\, GM/c^3$. Mass loading occurs at an average rate $\sim 10^{-2} \dot{M}$. 
Note that our definition produces a measurement that does not count mass that has been injected by the numerical floor prescription in the funnel: the tracer particles are initialized once, so the application of floors during the subsequent evolution does not increase the number of the tracer particles. We discard the beginning epoch of tracer data to avoid including the floors' effect on the transient tracer particle initial condition.

In both SANE and MAD models, mass entrainment is driven by instabilities in the boundary between the accreting plasma and the matter in the jet.  
Figure~\ref{fig:grmhd_projection} plots log plasma density on shells of constant radius over time and shows the development of an instability: as the high density midplane disk region moves to the right, it interacts with the low density funnel plasma moving to the left and forms Kelvin--Helmholtz rolls. Figure~\ref{fig:grmhd_equirectangular} plots density and specific angular momentum in the central frame of Figure~\ref{fig:grmhd_projection} in the $\theta-\phi$ plane at three different radii. Evidently, the KH roll is well resolved.

We observe that Kelvin--Helmholtz rolls develop in all simulations regardless of the accretion flow parameters; however, it is especially apparent in the MAD flows which have a more turbulent boundary layer. Mass entrainment thus proceeds in part through the Kelvin--Helmholtz instability at the jet--disk boundary. Still, the full structure of the jet--disk boundary layer is complicated, and braked accretion streams near the event horizon also contribute to mass loading.

\begin{figure*}[t]
    \centering
    \includegraphics[width=0.85\textwidth]{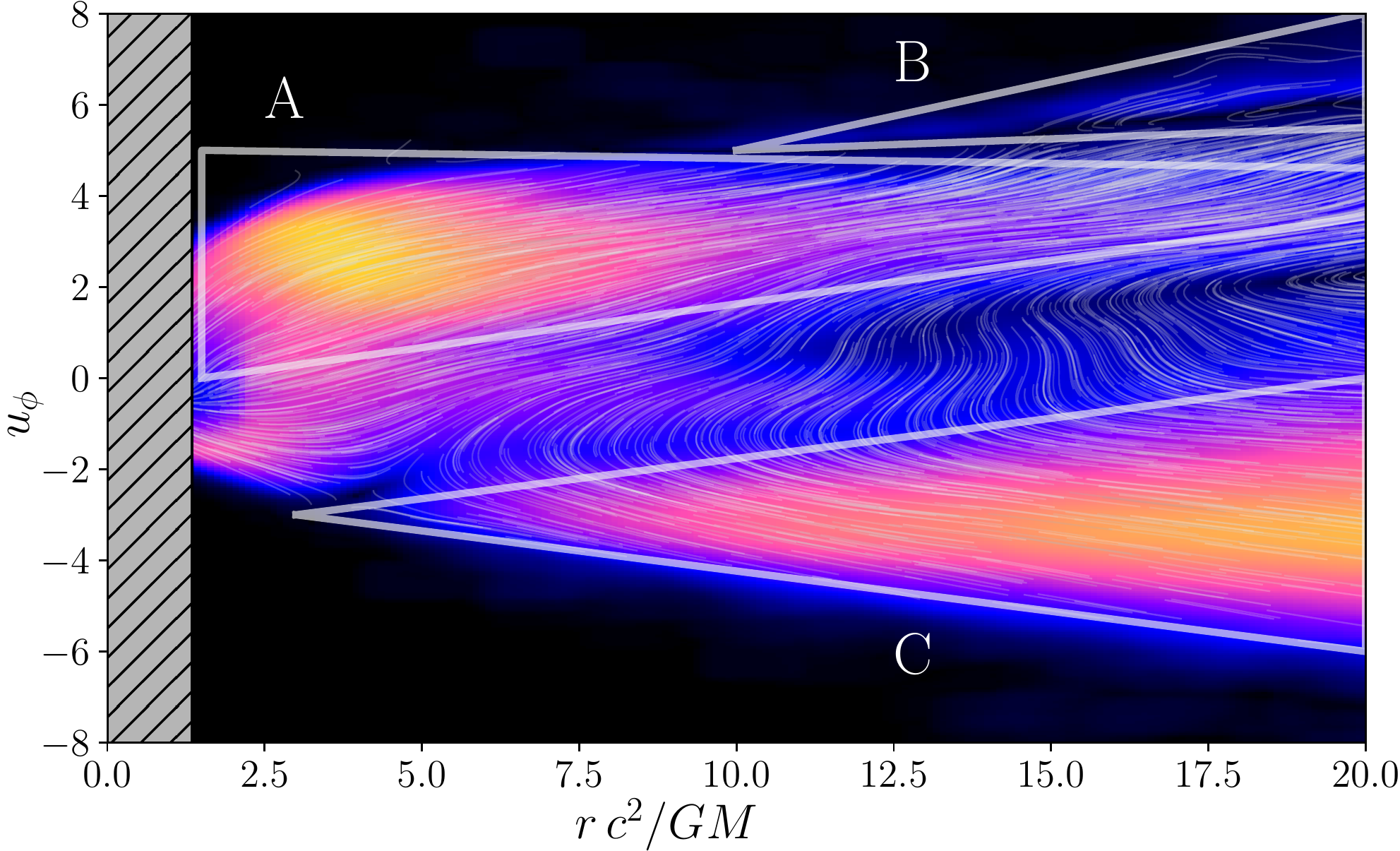}
    \caption{Time-averaged flow of tracer particles through the $r-u_\phi$ state space.  The gray hatched region at the left of the figure lies within the horizon. The background shows a false-color representation of the average speed of the particles through the two-dimensional state space and helps to visually differentiate the disk (region A), disk wind (region B), and jet (region C).
    The density of white lines is proportional to the density of particles in state space; for the purposes of visualization, the density is capped for regions in the disk that have large density. Average particle flow follows the thin white lines. As particles are entrained in the jet they cross $u_\phi=0$ and are then torqued and accelerate outwards. 
    }
    \label{fig:grmhd_tracer_stateimgs}
\end{figure*}

We also use the tracer particles to visualize the flow of matter through phase space. Figure~\ref{fig:grmhd_tracer_stateimgs} shows the time-averaged flow of tracer particles in the radius--specific angular momentum plane. Plasma density is represented by the density and thickness of the white flow lines.  Color denotes particle speed in phase space and helps differentiate between the disk/plunging region and the jet.

The flow at $r < 20$ can be divided into the three triangular regions shown in Figure \ref{fig:grmhd_tracer_stateimgs}. Region A contains particles that are falling towards the event horizon and gradually losing angular momentum.  It contains the plunging region (where the figure is brightest), the disk, and the characteristic MAD accretion streams seen in Figure~\ref{fig:grmhd_tracer_fallingin}. Region B is the disk wind. Region C is the jet. Particles enter the jet from Region A, are torqued until their angular momentum has the same sign as the black hole, and then are accelerated outward. Particles gain angular momentum as they accelerate away from the hole, as expected in a sub-Alfv\'enic wind.  

\section{Discussion}
\label{sec:discussion}

We have studied a set of retrograde MAD and SANE black hole accretion models.  We found that the angular momentum of plasma in both the jet and parts of the jet--disk boundary layer is aligned with the spin of the hole.
We also found that the boundary layer region, in which $u_\phi$ transitions between its value in the midplane and its value in the jet, was wider in the MAD models than in SANE models. This is unsurprising, since MAD flows tend to be more chaotic near the horizon where much of the jet--disk interaction occurs, so the time-averaged boundary location is spread out.
The existence of a shear layer is not restricted to retrograde models, as noted in \S\ref{sec:jet-disk-shear}, but we have focused on retrograde models because the shear is strongest there.

As noted in \S\ref{sec:mass-entrainment}, the jet--disk boundary is sufficiently resolved to see the development of Kelvin--Helmholtz rolls; this strongly suggests that numerical diffusion does not control the entrainment rate. Nevertheless increasing the simulation resolution may expose new structures, such as the plasmoids seen in recent high resolution axisymmetric models \citep{Nathanail2020, Ripperda2020}. 

To assess the effect of resolution we studied six different realizations of a MAD $\bhspin = -0.94$ model at four resolutions: two at 192 radial zones, one at 288, two at 384, and one at 448 (resolution in other coordinates is scaled proportionately).  We include multiple realizations at the same resolution to assess the error bars on measurements associated with turbulent fluctuations.  We consider convergence in two time-averaged quantities: the profiles of $u_\phi$ presented in Figure~\ref{fig:grmhd_funnel_profiles} and the total mass in the jet near the hole as measured from the GRMHD.

The time-averaged specific angular momentum profile $\left\langle u_\phi\right\rangle(r,\theta)$ is remarkably consistent across all resolutions everywhere except in the zones adjacent to the polar boundary, where we do not necessarily expect agreement because of our treatment of the boundary condition. In the shear region, the profiles are consistent to $5\%$ and exhibit no discernible trend with resolution. 

We compute the total mass in the jet near the hole by integrating the GRMHD density variable within a volume $V$
\begin{align}
    M_{\mathrm{j}}(t) \equiv \int\limits_{V} \rho \; \sqrt{-g} \, \ud r \, \ud \theta \, \ud \phi,
\end{align}
where we have chosen $V$ to be the region with $u_\phi < 0$ and $v^r > 0$ at $2 < r < r_* = 20$. Note that $M_{\mathrm{j}}(t)$ has contributions from both mass entrainment and numerical floors.
The time-dependent variation in the entrainment rate (see Figure~\ref{fig:grmhd_tracer_entrainment_times}), causes $M_{\mathrm{j}}(t)$ to fluctuate, so evaluations of the time-averaged $\langle M_{\mathrm{j}}(t)\rangle_t$ are subject to noise. We find that $M_{\mathrm{j}}(t)$ has a correlation time $\approx 200 \, GM/c^3$ in the MAD, $\bhspin = -0.94$ model.  The full model duration is $20,000 \, GM/c^3$, but the first $5,000 \, G M/c^3$ is an unequilibrated transient, so we have $N \sim 80$ independent samples over the full model; therefore, we expect fractional errors of order $N^{-1/2} \sim 10\%$. We find that $\langle M_{\mathrm{j}}(t)\rangle = 140, 130, 160$, and $130$ for simulations with radial resolution 192, 288, 384, and 448 respectively, which is consistent with the expected error. We also note that the widths of the jet and boundary-layer regions (in zones) reported in \S\ref{sec:funnel_profiles} scales linearly with the simulation resolution.

There may be additional mixing processes that occur on unresolved scales, so the consistency of $M_{\mathrm{j}}$ across resolutions does not prove that we have accurately accounted for mass mixing between the jet and disk.  Future convergence studies should probe not only longer timescales to reduce the fluctuation noise but also higher resolution. 

We also note that since the equilibration time increases with radius, the long-term average  jet--disk interaction may be poorly represented at large radii where the disk is still strongly dependent on initial conditions. We have chosen to overstep this issue by only reporting fits and statistics from equilibrated parts of our simulations. 
\citet{Chatterjee2019agnjet} also studied mass loading in their study of black hole jet launching. They performed multiple long-time, large-scale ($r_{\mathrm{max}} \gtrsim 10^5\;GM/c^2$) 2D GRMHD simulations and found that additional mass entrainment occurred at large radii. As noted above, the details of the jet--disk interaction at such large radii may be influenced by the choice of initial condition.

\section{Summary}
\label{sec:summary}

We have studied a set of three-dimensional GRMHD simulations of retrograde SANE and MAD black hole accretion disks at $\bhspin = -0.5$ and $-0.94$, with a focus on the jet--disk boundary near the horizon.  We have found that:

\vspace{0.5em}

1. Plasma in the jet rotates with the hole and not the disk. This generates a jet--disk boundary with strong currents and vorticity. \vspace{0.2em}

2. In MAD models accretion occurs through narrow plasma streams near the horizon.  These streams erode as they interact with the counterrotating jet, loading the jet with plasma. \vspace{0.2em}

3. In both MAD and SANE models, disk plasma is entrained in the jet in well-resolved Kelvin--Helmholtz rolls. \vspace{0.2em}

4. The entrainment rate is $\sim 0.01 \, \dot{M}$ for the MAD, $\bhspin = -0.94$ model that we are able to study in detail. \vspace{0.2em}

5. The entrainment rate and boundary layer structure are insensitive to resolution over the range in resolution we are able to study. \vspace{0.2em}

6. In retrograde MAD models accretion near the horizon fluctuates strongly: individual timeslices do not look like time- and azimuth- averaged data.  Relatedly, the jet in MAD models wobbles significantly. The fluctuations create a complicated interface between jet and disk.

\vspace{0.8em}

This study has considered a limited range of models and could be extended by comparing a broader range of black hole spins and tilts between the hole and the accretion flow. Understanding the behavior of jet plasma and the jet--disk boundary layer may be crucial in developing a robust model of the connection between black hole spin and motion in the jet, which can now be resolved in time and space by the Event Horizon Telescope.

\acknowledgements

The authors would like to thank the Event Horizon Telescope collaboration, especially Jason Dexter, Ramesh Narayan, and Andrew Chael, as well as Eliot Quataert and Patrick Mullen, for stimulating discussions.  The authors also thank Hector Olivares and the anonymous referee for their insightful comments and suggestions that improved the clarity of the manuscript.  The authors acknowledge the Texas Advanced Computing Center (TACC) at The University of Texas at Austin for providing HPC resources that have contributed to the research results reported within this paper.

This work was supported by NSF grants AST 17-16327 and OISE 17-43747.  GNW was supported in part by a Donald and Shirley Jones Fellowship and a research fellowship from the University of Illinois.  BSP was supported in part by the US Department of Energy through Los Alamos National Laboratory. Los Alamos National Laboratory is operated by Triad National Security, LLC, for the National Nuclear Security Administration of the US Department of Energy (Contract No.~89233218CNA000001).  CFG was supported in part by a Richard and Margaret Romano Professorial scholarship.

\software{NumPy \citep{numpy}, OpenCV \citep{opencv}, Matplotlib \citep{matplotlib}}

\pagebreak

\appendix

\section{FMKS Coordinates}
\label{sec:fmks}

The simulations in this paper were performed in funky modified Kerr--Schild (FMKS) coordinates $x^\mu = \left(x^0, x^1, x^2, x^3\right)$, which are an extension to the modified Kerr--Schild (MKS) coordinates introduced in \citet{Gammie2003}. Positive integer superscripts in this section should be interpreted as indices, not exponents. MKS coordinates are themselves a modification of the horizon-penetrating Kerr--Schild $x^{\overline{\mu}} = \left(t, r, \theta, \phi\right)$. Modifications were chosen to both reduce computational cost and increase effective resolution by concentrating zones in regions of the domain where more interesting physics occurs (like the midplane and near the horizon at small radii) and derefining unnecessary small zones. Each of FMKS, MKS, and KS is axisymmetric in $\phi$.

Both MKS and FMKS coordinates use an exponential radial coordinate $x^1 \equiv \log(r)$, which increases the number of zones at small radii where both the relevant dynamical timescale is shorter and it is more important to recover the detailed dynamics of the flow. 

FMKS makes two modifications to the elevation coordinate $x^2$. The first reproduces MKS and increases the number of zones near the midplane by introducing a sinusoidally varying dependence of $\Delta (x^2)$ on $\theta$, as
\begin{align}
    \theta_g \equiv \pi x^2 +  \dfrac{1}{2} \left(1 - h\right)\sin\left(2 \pi x^2\right)
\end{align}
where $h$ is the midplane finification parameter, which we set to $h = 0.3$. 

FMKS also introduces a cylindrification in $\theta$ whereby zones that are near the poles but are at small radii have larger elevational extent. This choice is meant to increase the required numerical timestep, which is set by the minimum of the signal-crossing time over all zones. The signal crossing time in zones near the funnel often approaches the speed of light, and thus this fact combined with the structure of spherical geometry (which keeps the number of azimuthal zones constant regardless of $\theta$) results in many small zones with fast signal speeds. Thus, through cylindrification, we increase the size of the smallest zones and similarly gain an increase in timestep. The cylindrification is achieved by defining 
\begin{align}
    \theta_j = N \left(2 x^2 - 1 \right) \left( 1 + \left(\dfrac{2 x^2 - 1}{B\left(1+\alpha\right)^{1/\alpha}}\right)^\alpha \right) + \pi / 2
\end{align}
where $\alpha$ and $B$ are parameters and where
\begin{align}
    N = \dfrac{\pi}{2} \left( 1 + \dfrac{B^{-\alpha}}{1 + \alpha} \right)^{-1}
\end{align} 
is a normalization term. Finally, the elevation coordinate is
\begin{align}
    \theta = \theta_g + \exp\left[ -s \Delta x^1 \right] \left( \theta_j - \theta_g \right)
\end{align}
where $\Delta x^1 = x^1 - \log\left[r_{\mathrm{in}}\right]$ measures the FMKS distance from the inner edge of the simulation. In our simulations, we take $s = 0.5, B= 0.82,$ and $\alpha = 14$.

We do not believe that the above coordinate definition is analytically invertible for $x^\mu(x^{\overline{\mu}})$. This is not a problem for codes that compute quantities numerically; however, for codes that require analytic forms of, e.g., the connection coefficients, these must either be computed beforehand otherwise a non-linear root finding step may be required to map KS locations into FMKS locations (e.g., if ray tracing).

\bibliography{main}
\bibliographystyle{aasjournal}
\end{document}